\documentclass[twocolumn,english,aps,prb,showpacs,byrevtex,amsmath,amssymb,superscriptaddress]{revtex4-1}
\usepackage{fancyhdr}
\usepackage{amsmath}
\usepackage{subfigure}
\usepackage{epsfig}
\usepackage{amssymb}
\usepackage{units} 
\usepackage{pifont}
\usepackage{textcomp}
\usepackage{gensymb}
\usepackage{pifont}
\usepackage{enumerate}
\usepackage{color}
\usepackage{float}
\usepackage{bm}
\usepackage{ulem}
\begin{document}

\title{Intrachain collinear magnetism and interchain magnetic phases \\ in Cr$_3$As$_3$-K-based materials}

\author{Giuseppe Cuono}
\affiliation{International Research Centre Magtop, Institute of Physics, Polish Academy of Sciences,
Aleja Lotnik\'ow 32/46, PL-02668 Warsaw, Poland}

\author{Filomena Forte}
\affiliation{Consiglio Nazionale delle Ricerche CNR-SPIN, UOS Salerno, I-84084 Fisciano (Salerno),
	Italy}
\affiliation{Dipartimento di Fisica "E.R. Caianiello", Universit\`a degli Studi di Salerno, I-84084 Fisciano
(SA), Italy}

\author{Alfonso Romano}
\affiliation{Dipartimento di Fisica "E.R. Caianiello", Universit\`a degli Studi di Salerno, I-84084 Fisciano
(SA), Italy}
\affiliation{Consiglio Nazionale delle Ricerche CNR-SPIN, UOS Salerno, I-84084 Fisciano (Salerno),
Italy}

\author{Xing Ming}
\affiliation{College of Science, Guilin University of Technology, Guilin 541004, PR China}

\author{Jianlin Luo}
\affiliation{Beijing National Laboratory for Condensed Matter Physics and Institute of Physics, Chinese Academy of Sciences,
Beijing 100190, China}
\affiliation{Songshan Lake Materials Laboratory, Dongguan, Guangdong 523808, China}
\affiliation{School of Physical Sciences, University of Chinese Academy of Sciences, Beijing 100190, China}

\author{Carmine Autieri}
\email{autieri@magtop.ifpan.edu.pl}
\affiliation{International Research Centre Magtop, Institute of Physics, Polish Academy of Sciences,
	Aleja Lotnik\'ow 32/46, PL-02668 Warsaw, Poland}
\affiliation{Consiglio Nazionale delle Ricerche CNR-SPIN, UOS Salerno, I-84084 Fisciano (Salerno),
	Italy}

\author{Canio Noce}
\affiliation{Dipartimento di Fisica "E.R. Caianiello", Universit\`a degli Studi di Salerno, I-84084 Fisciano
(SA), Italy}
\affiliation{Consiglio Nazionale delle Ricerche CNR-SPIN, UOS Salerno, I-84084 Fisciano (Salerno),
Italy}

\date{\today}
\begin{abstract}
We perform a comparative study of the KCr$_3$As$_3$ and the K$_2$Cr$_3$As$_3$ quasi 1D compounds, and show that
the strong interplay between the lattice and the spin degrees of freedom promotes a new collinear ferrimagnetic ground state within the chains in presence of intrachain antiferromagnetic couplings.
We propose that the interchain antiferromagnetic coupling in KCr$_3$As$_3$ plays a crucial role for the experimentally observed spin-glass phase with low critical temperature.
In the same region of the parameter space, we predict K$_2$Cr$_3$As$_3$ to be non-magnetic but on the verge of the magnetism, sustaining interchain ferromagnetic spin fluctuations while the intrachain spin fluctuations are antiferromagnetic.
\end{abstract}


\maketitle

\section{Introduction}


Recently, bulk superconductivity in K$_{2}$Cr$_{3}$As$_{3}$ with $T_{c}$=6.1 K has been reported, this representing the first observation of superconductivity in Cr-based compounds at ambient pressure  \cite{Bao15}. Following this discovery, three additional superconductors in the series A$_{2}$Cr$_{3}$As$_{3}$, with A=Na \cite{Mu18}, Rb \cite{Tang15} and Cs \cite{Tang15b}, have been grown up, with superconducting critical temperatures equal to 8.6 K, 4.8 K and 2.2 K, respectively. 
This is a new family of superconductors, likely to be unconventional  \cite{Goll06,Norman11,Zhi15,Adroja15,Pang15,Noce20,Zhong15}, which differently from the previously discovered Cr-based superconductor CrAs \cite{Chen19,Wu10,Wu14,Kotegawa14,Autieri17,Autieri17b,Autieri18,Cuono19,Cuono19b,Nigro19}, exhibits a quasi-one dimensional crystal structure, with infinite {[}(Cr$_{3}$As$_{3}$)$^{2-}${]}$_{\infty}$ linear
chains of double-walled sub-nanotubes (DWSN), interconnected
by A$^{+}$ cations \cite{Bao15, Mu18, Tang15, Tang15b}.
In particular, the role played in these compounds by the reduced dimensionality in conjunction with the electronic correlations is currently the subject of intense investigation \cite{Kong15,Watson17,Cao17,Cao18,Cuono18,Cuono19c,Reja19}.
We also point out interesting analogies between the properties of the above-mentioned one-dimensional Cr-based superconductors and the family of MQ$_2$ (M = Nb, Ta; Q = S, Se) superconductors with one-dimensional spectrum \cite{Wang20}  and star-David clusters \cite{Sipos08,Martino20,Wickramaratne20}.
\\

\begin{figure}[]
	\centering
	\includegraphics[width=\columnwidth, angle=0]{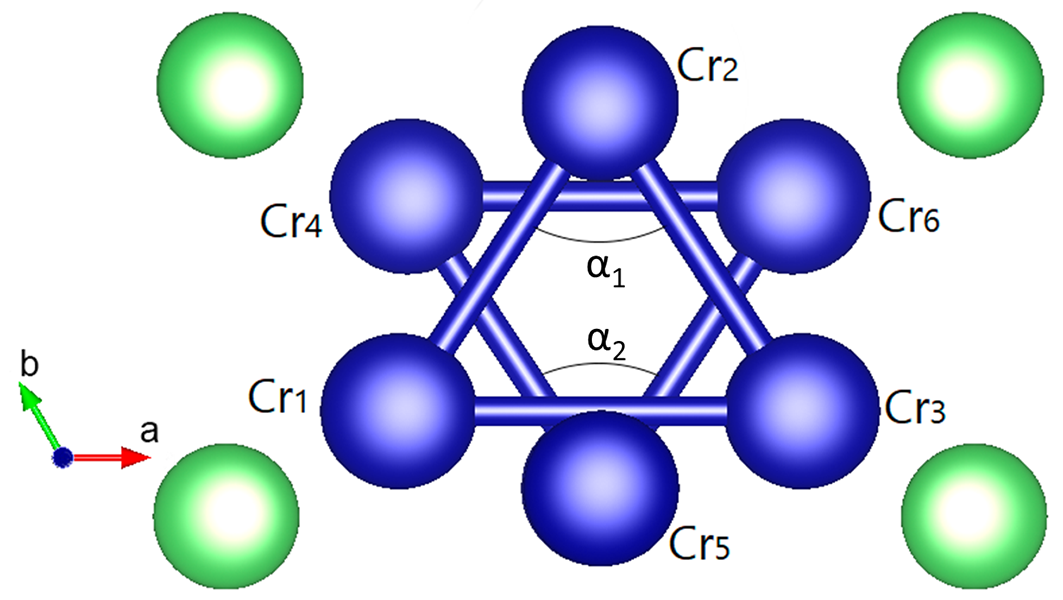}
	\caption{Cr-triangles belonging to the {[}(Cr$_{3}$As$_{3}$)$^{2-}${]}$_{\infty}$ subnanotubes, in the distorted case. Blue and green spheres denote Cr and As atoms, respectively.
	}
	\label{Triangle}
\end{figure}

Almost at the same time, a parent compound series has been synthesized, namely ACr$_{3}$As$_{3}$ (A = K, Rb, Cs), which, according to some authors does not exhibit superconductivity but a spin glass-like magnetism \cite{Bao15b}, while according to other authors presents bulk superconductivity \cite{Mu17,Liu18}.
It is argued that superconductivity arises both in KCr$_{3}$As$_{3}$ \cite{Mu17} and RbCr$_{3}$As$_{3}$ \cite{Liu18} upon post-treatment with a hydrothermal annealing in pure dehydrated ethanol, this leading
to an improvement of the sample crystallinity.
Recently, Taddei et al. \cite{Taddei19} detected a previously missed effect of the ethanol bath alkaline ion deintercalation. They suggest that the difference between non-superconducting spin-glass and non-magnetic superconducting samples is not related to the sample crystallinity but rather to the amount of intercalated hydrogen\cite{Taddei19,Wu19}. In this context, a more effective way of intercalating hydrogen in KCr$_{3}$As$_{3}$ has very recently been realized via electrochemical hydrogenization in KOH solution at about   $75^\circ$C. This procedure led to the fabrication of samples which exhibit a novel phase separation regime where a minority superconducting phase with $T_c$ as high as 5.8\,K coexists with a majority nonsuperconducting one \cite{Xiang20}. 
\\

Therefore, both the A$_{2}$Cr$_{3}$As$_{3}$ and the ACr$_{3}$As$_{3}$ families still need a deep analysis to better understand the interplay between structural, magnetic and superconducting properties.
\\

As far as magnetism is concerned, in the case of K$_{2}$Cr$_{3}$As$_{3}$ first-principles calculations \cite{Jiang15} suggest that the triangular geometry tends to frustrate antiferromagnetism, so that the nonmagnetic phase is the most stable one. On the other hand, Wu et al. \cite{Wu15} predict that K$_{2}$Cr$_{3}$As$_{3}$ and Rb$_{2}$Cr$_{3}$As$_{3}$ possess strong frustrated magnetic fluctuations and are near a novel in-out co-planar magnetic ground state. Interestingly, magnetism increases in the A$_{2}$Cr$_{3}$As$_{3}$ compounds when we go from the K to the heavier Cs. 
Nuclear quadrupole resonance measurements on the A$_{2}$Cr$_{3}$As$_{3}$ family indicate that going along the series A=Na, Na$_{0.75}$K$_{0.25}$, K, Rb the system tends to approach a possible ferromagnetic quantum critical point \cite{Luo19}.
For KCr$_{3}$As$_{3}$, Bao et al. \cite{Bao15b} report that the effective magnetic moment is 0.68 $\mu_{B}$/Cr, with a susceptibility that below 56 K exhibits a behavior deviating from the standard Curie-Weiss one. In the same temperature range a rapid increase of the resistivity is also found, this suggesting the formation of spin clusters. Moments then freeze into a spin-glass state below  $T_{f}$= 5 K, as also signalled by the behavior of the specific heat which, as in other spin glass systems, shows a peak slightly above $T_{f}$  \cite{Bao15b}. These results seem to be related to the geometrical frustration between the Cr local spins, though the microscopic origin of the spin-glass phase is not addressed.

Very recently, neutron total scattering and density functional theory (DFT) studies \cite{Taddei18} have revealed significant phonon instabilities, associated to a frustrated orthorhombic distortion in K$_{2}$Cr$_{3}$As$_{3}$. Large atomic displacement parameters with anomalous temperature dependencies have been found, which result from highly localized orthorhombic distortions of the CrAs sublattice and coupled K displacements \cite{Taddei18}. The Cr-triangles in the double walled subnanotubes are no longer equilateral, which could lead to the release of magnetic frustration. These results suggest a more complex phase diagram with a subtle interplay of structural, electron-phonon and magnetic interactions. Further investigation is thus needed for K$_{2}$Cr$_{3}$As$_{3}$ as well as for other superconductors belonging to the same class, with the aim of understanding to what extent their superconducting behavior should be considered unconventional.
A lattice instability has also been found in KCr$_{3}$As$_{3}$, corresponding to a distortion of the Cr metallic wires in the crystal structure. This distortion couples strongly to both the electronic
and the magnetic properties \cite{Xing19, Xu20}.

In this paper we address how the magnetic properties of K$_{2}$Cr$_{3}$As$_{3}$ and KCr$_{3}$As$_{3}$ change by taking into account the distortions predicted in recent literature \cite{Taddei18}. We investigate the magnetism in both compounds, in the case where deformations of the chromium ion triangles consistent with the orthorhombic distortions are considered. 
We use a DFT approach, also including the Coulomb repulsion $U$, to explore the most favorable magnetic configurations inside the chains and between the chains.  We get the optimized crystal structure and show that the strong interplay between the lattice and the spin degrees of freedom promotes a new collinear ferrimagnetic ground state
within the chains. We propose that the experimentally observed spin glass phase at low temperature in KCr$_{3}$As$_{3}$ can be attributed to geometric frustration of antiferromagnetic coupling among the chains. 
Moreover, we show that the K$_{2}$Cr$_{3}$As$_{3}$ non-magnetic state is in close proximity to a ferrimagnetic phase, due to the ferromagnetic interchain magnetic exchange emerging in a regime of moderate electronic correlations.

The paper is organized as follows: in Sec.~II we describe in detail the crystal structures of K$_{2}$Cr$_{3}$As$_{3}$ and KCr$_{3}$As$_{3}$, in Sec.~III we report the computational details of our approach,
Sec.~IV is devoted to the magnetic properties inside the subnanotubes, in Sec.~V we introduce an Heisenberg model for the magnetic exchanges inside the chain, in Sec.~VI we present a comparative analysis between K$_{2}$Cr$_{3}$As$_{3}$ and KCr$_{3}$As$_{3}$, showing the results emerging when the magnetic interactions between the chains are considered, and finally last Section is devoted to a summary discussion and to the conclusions.

\section{Crystal structure of C\lowercase{r}$_3$A\lowercase{s}$_3$-chain-K-based materials}

The crystal structure of K$_2$Cr$_3$As$_3$ is  quasi-one-dimensional, according to the needle-like
morphology experimentally verified using single-crystal X-ray diffractions \cite{Bao15}. The structure contains infinite [(Cr$_3$As$_3$)$^{2-}$]$_{\infty}$ linear chains of DWSN, interconnected by K$^+$ cations (see Fig.~\ref{Triangle}). Cr atoms should bond covalently with As, whereas As should bond ionically with K$^+$, separating electro-positive Cr and K atoms. The [(Cr$_3$As$_3$)$^{2-}$]$_{\infty}$  DWSN are composed of inner Cr$_3$ twisted tubes and outer As$_3$ ones, which are constructed by the face-sharing Cr$_6$ (or As$_6$)
octahedra along the crystallographic $c$ direction \cite{Bao15}. Since the Cr sublattices may carry magnetic moments, one expects strong geometric magnetic frustration.

The [(Cr$_3$As$_3$)$^{2-}$]$_{\infty}$ DWSN and the K$^+$ cations form a hexagonal lattice with the space group of P$\bar{6}$m2. Every unit cell contains two formula units, the
chemical formula of one unit cell being K$_4$Cr$_6$As$_6$. We note that all the atoms occupy the crystalline planes $z=0$ and $z=0.5$, with the two crystallographically different K sites, namely K$_1$ and K$_2$, located at $z=0.5$ and $z=0$, respectively. This arrangement 
leads to the absence of inversion symmetry as well as to the loss of six-fold rotation symmetry.
Correspondingly, there exist two inequivalent As and Cr sites. 

\begin{figure*}
  \includegraphics[scale=0.35]{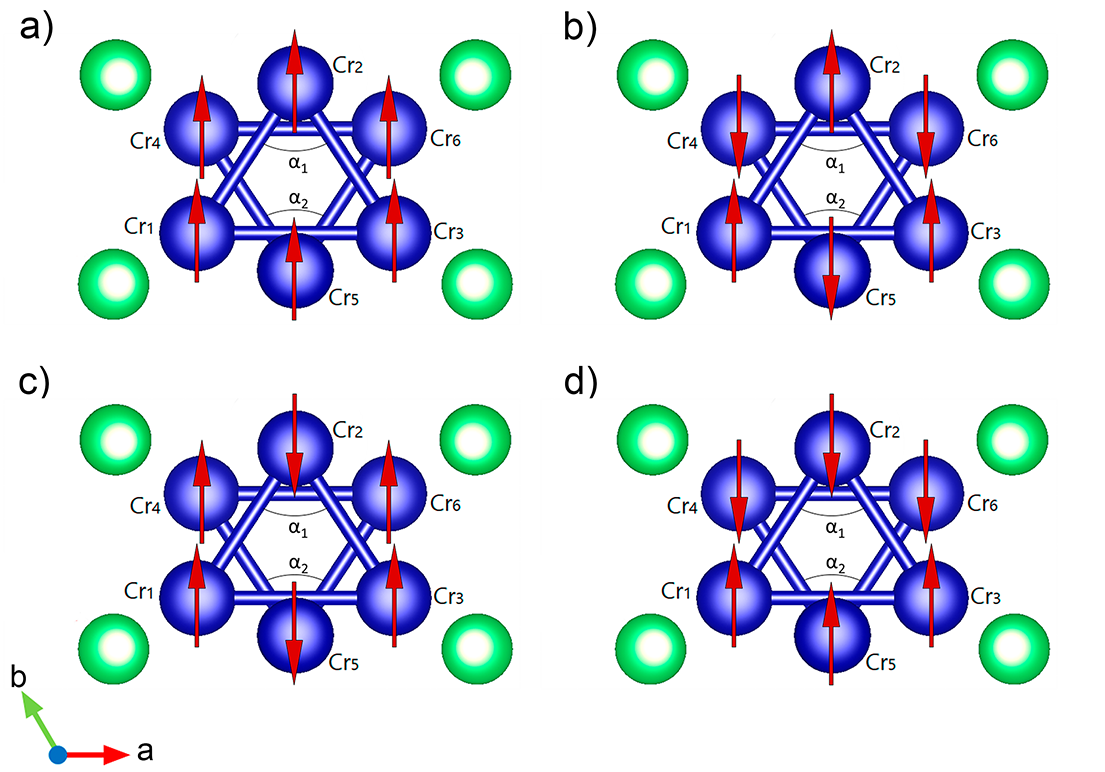}
	\caption{Cr-spin configurations investigated in the paper. We have defined these states as: a) the ferromagnetic state (FM), b) the interlayer antiferromagnetic state  (AFM), c) the up-up-down/up-up-down  ($\uparrow\uparrow\downarrow$-$\uparrow\uparrow\downarrow$) stripe state, d) the up-up-down/down-down-up  ($\uparrow\uparrow\downarrow$-$\downarrow\downarrow\uparrow$) zig-zag state.
	}
	\label{configurations}
\end{figure*}

The K$_2$Cr$_3$As$_3$ material is not stable since chemically it  deteriorates easily at ambient
conditions. However, by means of  a topotactic reaction, that keeps the Cr$_3$As$_3$ chains unmodified, one gets the corresponding parent compound KCr$_3$As$_3$, which, on the contrary, is
stable in air. This compound loses two K ions in a unit cell, and correspondingly
the lattice parameters $a$ and $c$ of KCr$_3$As$_3$ decrease by
9$\%$ and 1$\%$, respectively, compared with those of K$_2$Cr$_3$As$_3$. More importantly, there is only one site for K ions, which changes the point group from D$_{3h}$ to C$_{6h}$, and the space group
from P$\bar{6}$m2 (No. 187) \cite{Bao15} to P6$_3$/m (No. 176) \cite{Bao15b}. Owing to the symmetry, all the Cr triangles in the $ab$ plane have exactly the same size. 
Thus, we may guess that the different space group could affect in a distinct way the electronic and magnetic properties of the two different K-based materials.

\section{Approach and computational details}

We have performed DFT calculations by using the VASP package \cite{Kresse93,Kresse96,Kresse96b}. 
The core and the valence electrons were treated within 
the Projector Augmented Wave (PAW) \cite{Kresse99} method with a cutoff of 440 eV for the plane wave basis.
For calculations concerning the interactions inside a single chain, we have used the PBEsol exchange-correlation method \cite{Perdew08}, 
a revised Perdew-Burke-Ernzerhof (PBE) Generalized Gradient Approximation (GGA) that improves equilibrium properties of solids. 
This choice is motivated by the fact that the functional used within PBEsol is the most suited for the relaxation and in this system 
the relaxation is fundamental.
These calculations have been performed using a 4$\times$4$\times$10
$k$-point grid, in such a way to have 160 $k$-points in the first Brillouin zone. 

In DFT, electronic interaction energies are simply
described as the sum of the classical Coulomb repulsion between electronic densities in a mean field
approximation, via Hartree term, and the exchange-correlation term that is supposed to encompass all the correlations
and spin interactions. This approach has proved to be quite efficient in studies of weakly correlated materials.  
When moderately and strongly correlated electron systems are instead considered, an additional Coulomb repulsion $U$ is added to the energy functional\cite{Liechtenstein95}.
The LDA+U increases the magnetic moment with respect to LDA. In metallic systems, large values of U may give rise to an overestimation of the magnetic moment, so that in the case of moderate correlations quite small values of the Coulomb repulsion are usually required to reproduce the experimental results. We also remind that in metallic compounds the Fermi screening is expected to reduce the electrostatic repulsion.

PBEsol + $U$ is actually the approach that we have followed here to take into account the correlations associated with the Cr 3$d$ states. 
The values of $U$ specifically suited for the investigation of K$_2$Cr$_3$As$_3$ and KCr$_3$As$_3$ have been selected on the basis of a scanning of the magnetic moment as a function of the Coulomb repulsion. Starting from the fact that for $U=0$ we always get a vanishing magnetic moment, we consider gradually increasing values of $U$, in this way reproducing the entire spectrum of the magnetic moment values.
Comparing the obtained results with the experiments, we came to the conclusion that within our theoretical setup a suitable value of the effective Coulomb repulsion is $U=0.3\,$eV for both compounds.
 
In the numerical procedure we have relaxed the lattice constants and the internal atomic positions, with the forces that have been minimized to less than 0.01 eV/{\AA} in the structural relaxation.
To avoid adding an additional degree of freedom, we have fixed the volume of the unit cell to its experimental value and then performed the relaxation of the atomic positions.
For the calculations between different chains, we have used the PBEsol, the local density approximation (LDA) and the SCAN METAGGA \cite{Sun15}.
These calculations have been performed using a 2$\times$4$\times$10
$k$-point grid, and thus 80 $k$-points in the first Brillouin zone; we have halved the number of $k$ points along the $x$ direction because the number of cells along $x$ has doubled.
The values of the lattice constants of K$_{2}$Cr$_{3}$As$_{3}$ are $a$= 9.9832 {\AA} and $c$= 4.2304 {\AA} \cite{Bao15} and of KCr$_{3}$As$_{3}$ are $a$= 9.0909 {\AA} and $c$= 4.1806 {\AA} \cite{Bao15b}.

\section{Ab-initio magnetic properties of {[}(Cr$_{3}$As$_{3}$)$^{2-}${]}$_{\infty}$ subnanotubes}
It is well-established that the significant phonon instability found in K$_2$Cr$_3$As$_3$ may give rise to a frustrated orthorhombic distortion\cite{Taddei18}. This implies that the
Cr triangles in the DWSN are no longer equilateral, likely leading to a subtle interplay between the magnetic frustration and the structural properties of the material.

\begin{figure}[]
	\centering
	\includegraphics[width=\columnwidth, angle=0]{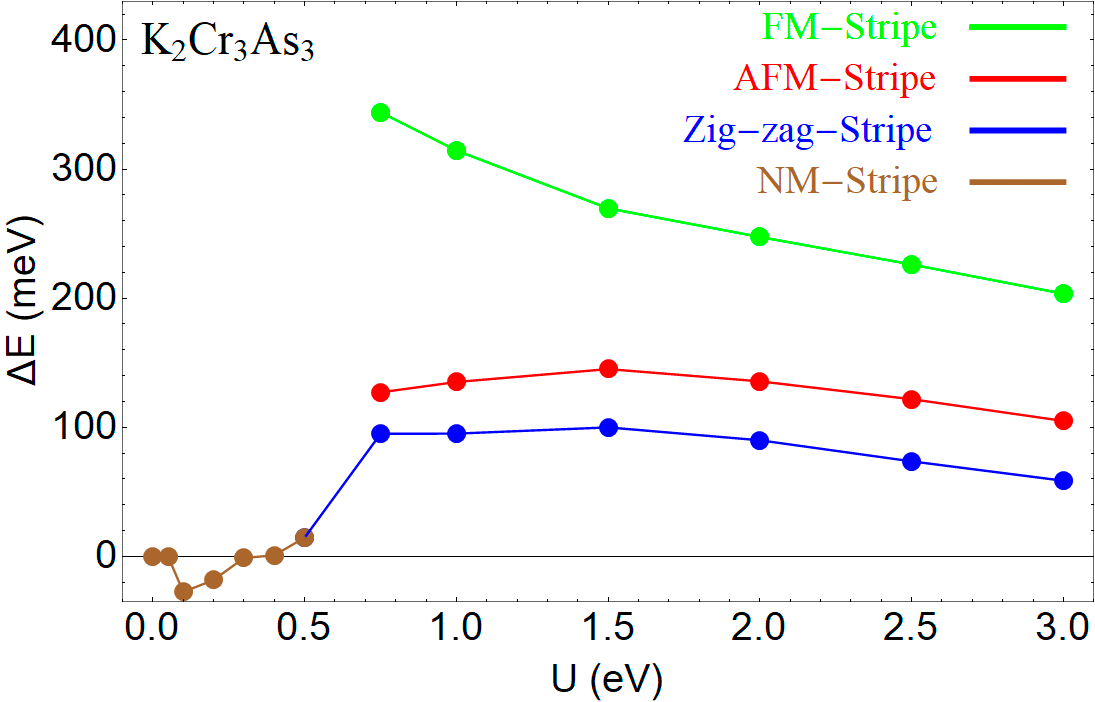}
	\caption{ Energy per Cr-atom of the FM, AFM, zig-zag and NM states measured with respect to the stripe state as a function of the Coulomb interaction for the K$_{2}$Cr$_{3}$As$_{3}$ compound. At low values of $U$ data for the FM and the AFM phases are not shown due to the lack of convergence of the numerical procedure in that regime.} 
	\label{EC_K}
\end{figure}

To investigate this issue we have performed the atomic relaxation of the distorted triangles composed of Cr-atoms and belonging to the {[}(Cr$_{3}$As$_{3}$)$^{2-}${]}$_{\infty}$ subnanotubes, for different values of the Coulomb repulsion $U$.
The relaxation procedure has been done starting from both collinear and non-collinear magnetic configuration. In the undistorted phase, assuming an initial non-collinear magnetic configuration, convergence is reached maintaining this kind of configuration. Performing the same procedure in the distorted phase, the system converges instead to a collinear magnetic configuration, thus implying that in this case the non-collinear solution is metastable. 
Therefore, the attainable magnetic configurations here considered are planar-collinear arrangements of the Cr magnetic moments within the unit cell in the DWSN. Our DFT analysis shows that in presence of $U$ these collinear phases are the magnetic stable states, instead of the non-collinear ones previously predicted \cite{Jiang15,Wu15}. They are reported in Fig.~\ref{configurations}, and correspond to the ferromagnetic state (FM), the interlayer antiferromagnetic state  (AFM), the up-up-down/up-up-down stripe state ($\uparrow\uparrow\downarrow$-$\uparrow\uparrow\downarrow$) and the up-up-down/down-down-up zig-zag state ($\uparrow\uparrow\downarrow$-$\downarrow\downarrow\uparrow$). Together with these, we also consider the non-magnetic (NM) one. We point out that the notation adopted for the atoms and the angles of the Cr triangles is the same as the one reported in Fig.~\ref{Triangle}. Thus, the configurations considered in our calculations are represented in Fig.~\ref{configurations}. In this Figure Cr$_1$, Cr$_2$ and Cr$_3$ indicate the Cr atoms belonging to $z=0$ plane, whereas  Cr$_4$, Cr$_5$ and Cr$_6$ are the Cr atoms in the $z=0.5$ plane. Moreover, the angle $\alpha_1$ ($\alpha_2$) is the angle at the vertex of the triangle of the plane $z=0$ ($z=0.5$). 
In the distorted case the Cr-atoms at different planes are closer than in the undistorted case, so that, for instance, the Cr$_2$-Cr$_4$ and the Cr$_2$-Cr$_6$ distances become shorter. Therefore, the coupling between the atoms at $z=0$ and $z=0.5$ is stronger, affecting in a different way the electronic properties as well as the magnetic couplings, as shown in Appendices A and B.
In the next subsections we will present our numerical DFT calculations for the two {[}(Cr$_{3}$As$_{3}$)$^{2-}${]}$_{\infty}$ DWSN in K$_{2}$Cr$_{3}$As$_{3}$ and KCr$_{3}$As$_{3}$, respectively.

\subsection{K$_2$Cr$_3$As$_3$}
In Fig.~\ref{EC_K} we plot the energy of the above mentioned configurations as a function of the Coulomb repulsion $U$, evaluated with respect to the energy of the stripe state. We observe that for values of the Coulomb repulsion $U<U_c=0.4\,$eV, the ground state is non-magnetic, then becoming the stripe one for $U\geq U_c$. We also note that the energy of the ferromagnetic configuration is always larger than the other collinear considered phases. 

\begin{figure}[]
	\centering
	\includegraphics[width=\columnwidth, angle=0]{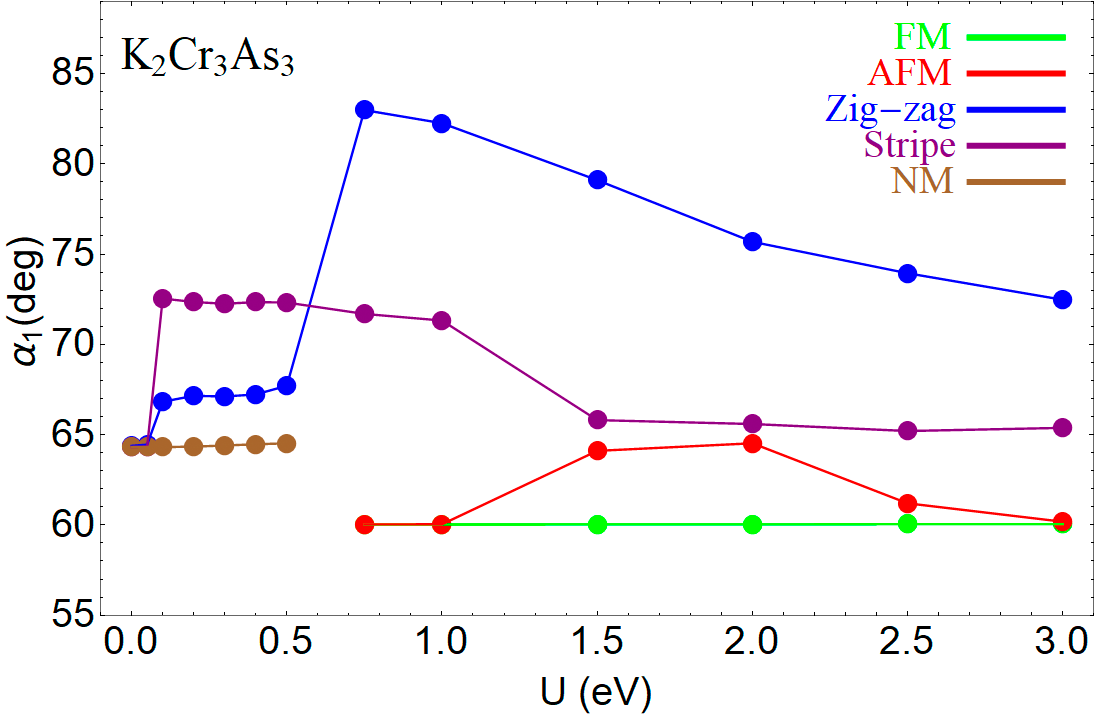}
	\includegraphics[width=\columnwidth, angle=0]{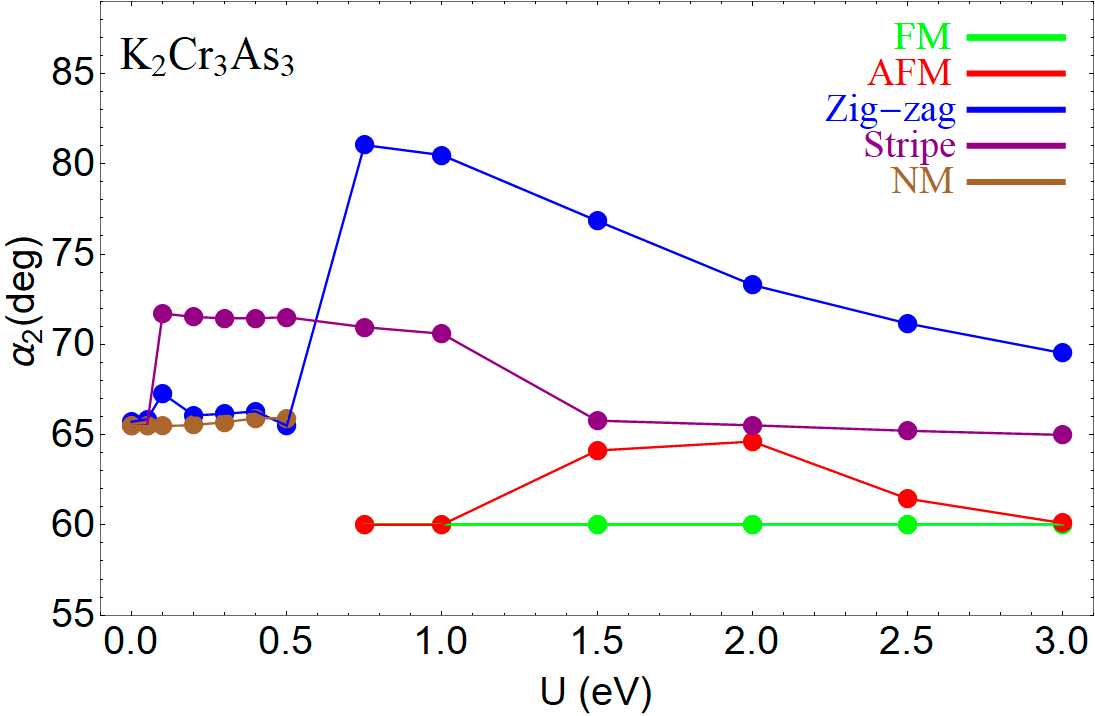}
	\caption{Angles of the Cr-triangles in K$_{2}$Cr$_{3}$As$_{3}$ as functions of the Coulomb interaction. In the upper panel $\alpha_1$ is the angle at the vertex of the Cr-triangle located at $z$=0, while in the lower panel $\alpha_2$ is the same for the triangle located at $z$=0.5.
	}
	\label{angles_K}
\end{figure}

\begin{figure*}[]
	\centering
	\includegraphics[width=\columnwidth,  angle=0]{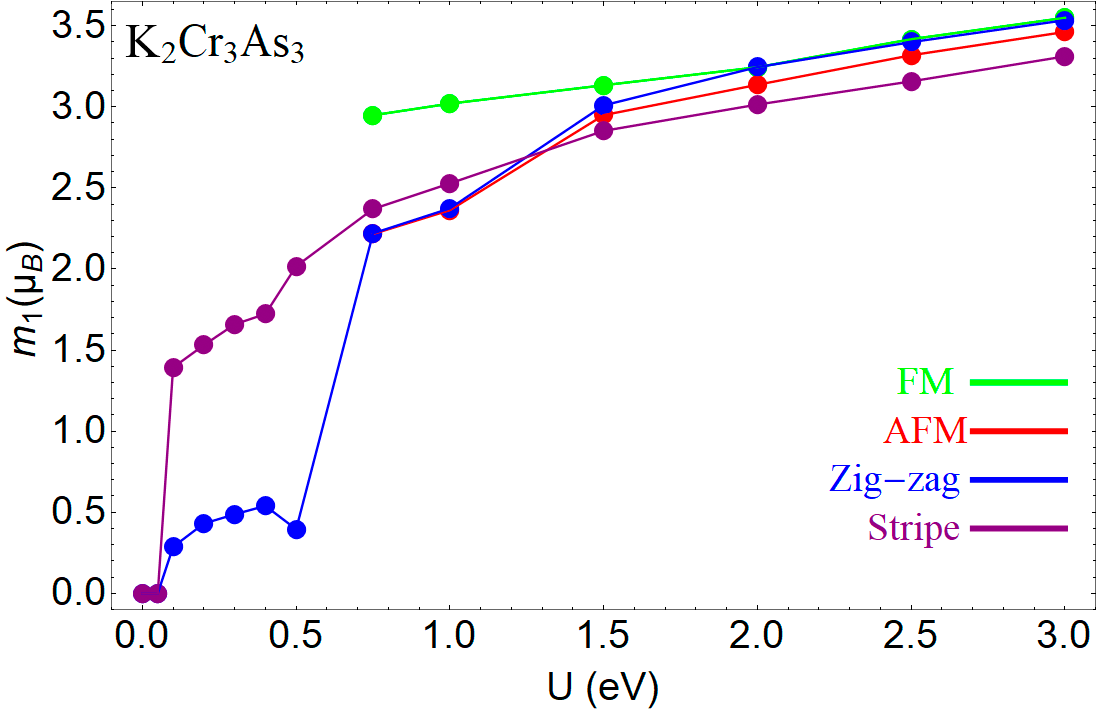}
	\includegraphics[width=\columnwidth,  angle=0]{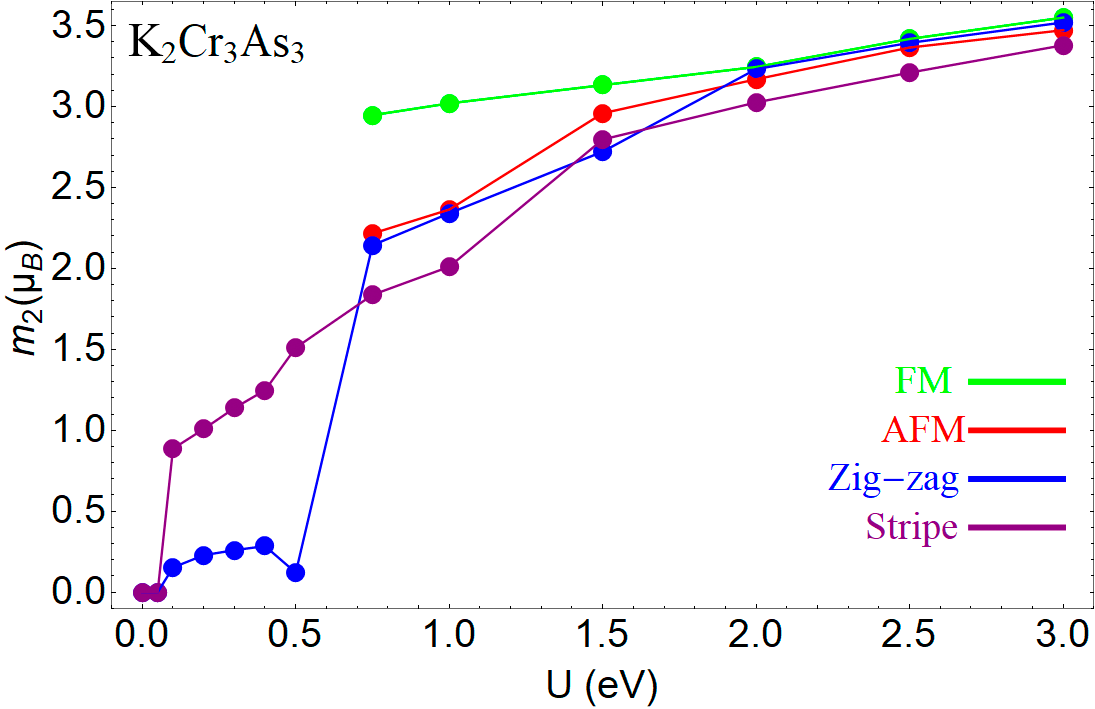}
	\includegraphics[width=\columnwidth, angle=0]{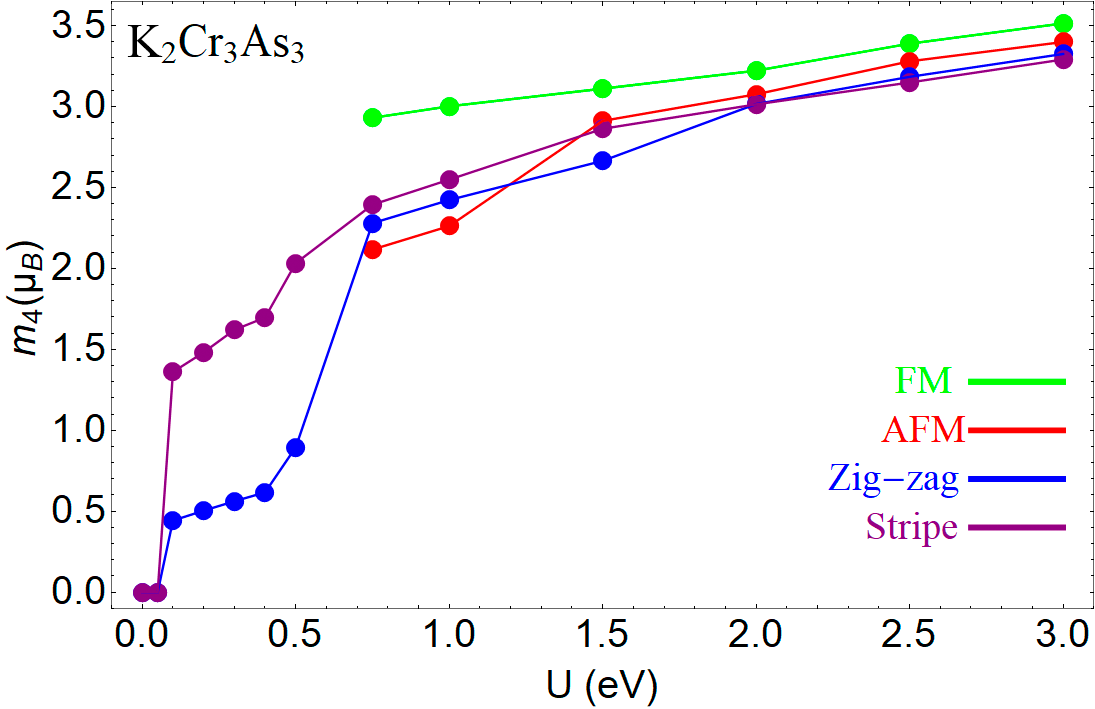}
	\includegraphics[width=\columnwidth, angle=0]{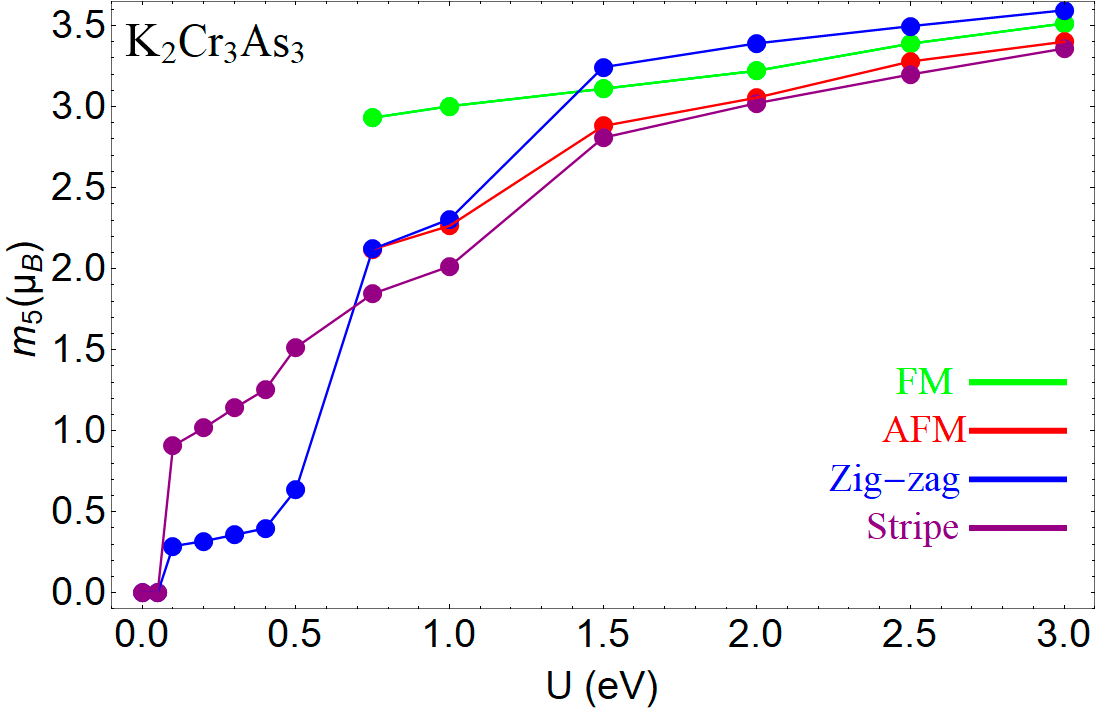}
	\caption{Magnetic moments of the Cr ions for the configurations shown in Fig. \ref{configurations}, as functions of the Coulomb interaction for the K$_{2}$Cr$_{3}$As$_{3}$ compound.}
	\label{magneticmoment_K}
\end{figure*}

Figs.~\ref{angles_K} and \ref{magneticmoment_K} report the behavior as functions of $U$ of the angles $\alpha_1$ and $\alpha_2$ shown in Fig.~\ref{configurations} and of the four inequivalent magnetic moments of the Cr ions, respectively.
From these figures, we deduce that the system is characterized by a strong interplay between lattice and spin degrees of freedom, implying that it is not possible to decouple the magnetic properties from the structural ones. Indeed, we can see from Fig.~\ref{angles_K} that, apart from the FM configuration, the other possible magnetic states always correspond to distorted triangles, with a distortion which for $U$ larger than 0.5\,eV is particularly pronounced in the zig-zag phase. We also note that above $U=1.5\,$eV there are no more structural effects on the stripe phase, whereas for the other magnetic configurations there is still a dependence of $\alpha_1$ and $\alpha_2$ on the Coulomb repulsion. Importantly, this trend is quite similar for the triangles located at $z=0$ and $z=0.5$. When lower values of $U$ are considered, we observe that the lowest energy NM and stripe phases correspond to distorted triangles, with the angles $\alpha_1$ and $\alpha_2$ staying unaffected upon variations of the Coulomb repulsion $U$.

\begin{figure}[]
	\centering
	\includegraphics[width=\columnwidth, angle=0]{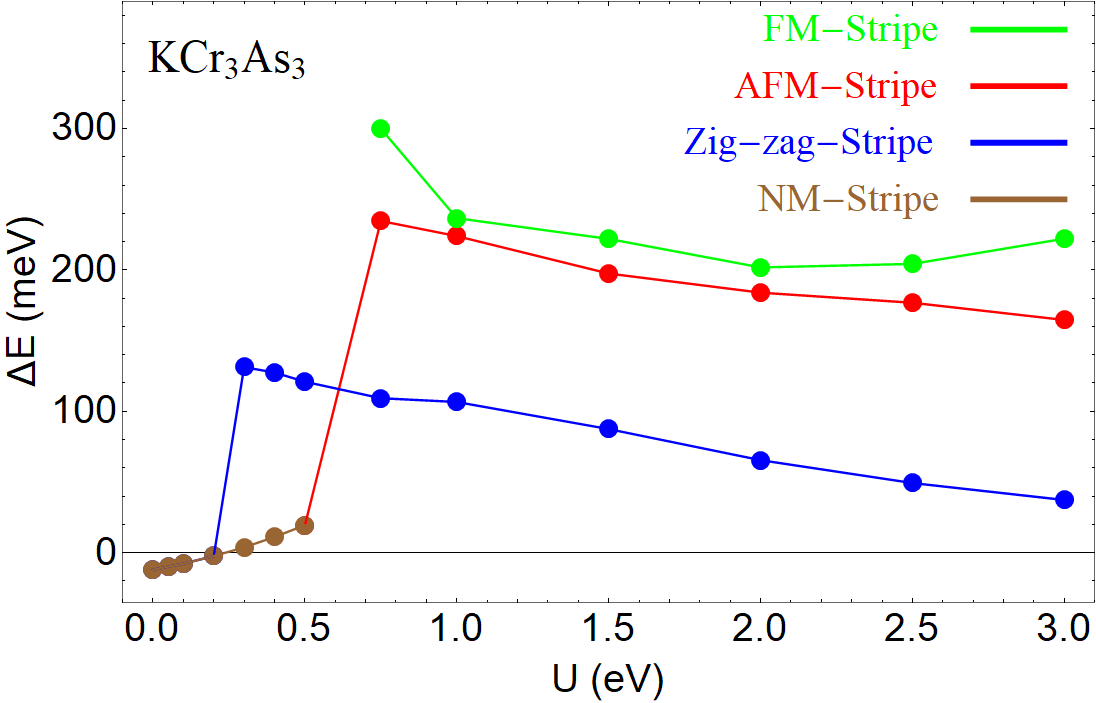}
	\caption{Energy per Cr-atom of the FM, AFM, zig-zag and NM states measured with respect to the stripe state, as a function of the Coulomb interaction for the KCr$_{3}$As$_{3}$ compound.}
	\label{EC_KCr3As3}
\end{figure}

\begin{figure}
	\centering
	\includegraphics[width=\columnwidth, angle=0]{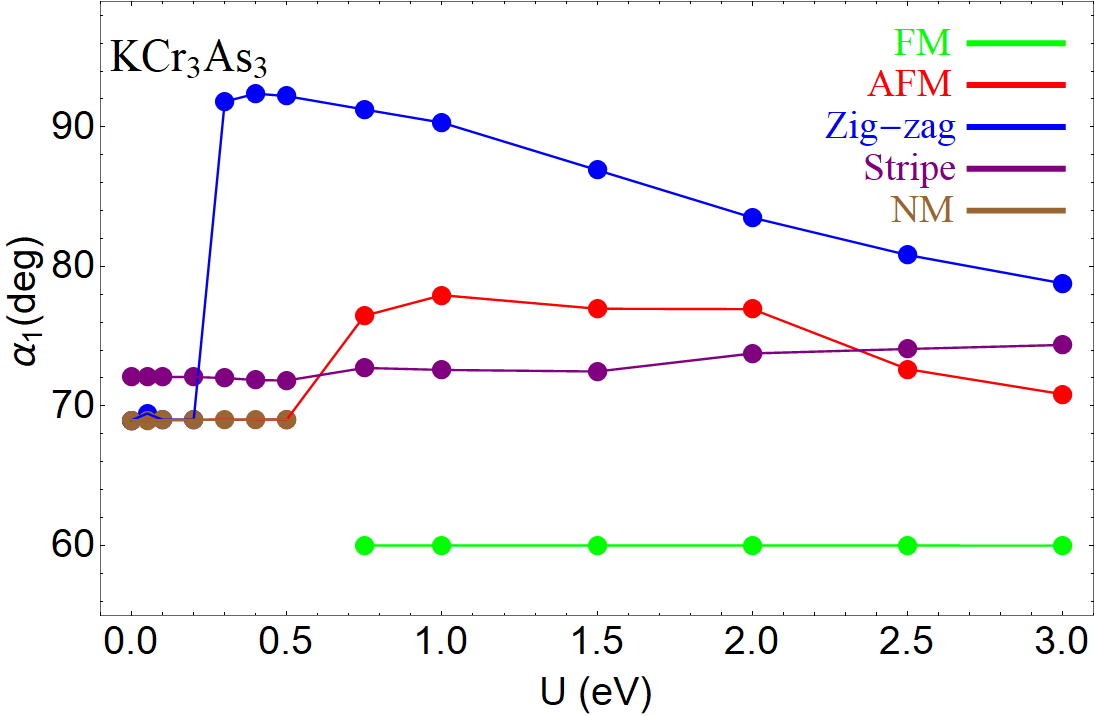}
	\caption{Angle at the vertex of the Cr-atom triangle at $z$=0 as a function of the Coulomb interaction for the KCr$_{3}$As$_{3}$ compound.
	}
	\label{angles_KC3As3}
\end{figure}

\begin{figure*}[]
	\centering
	\includegraphics[width=\columnwidth,  angle=0]{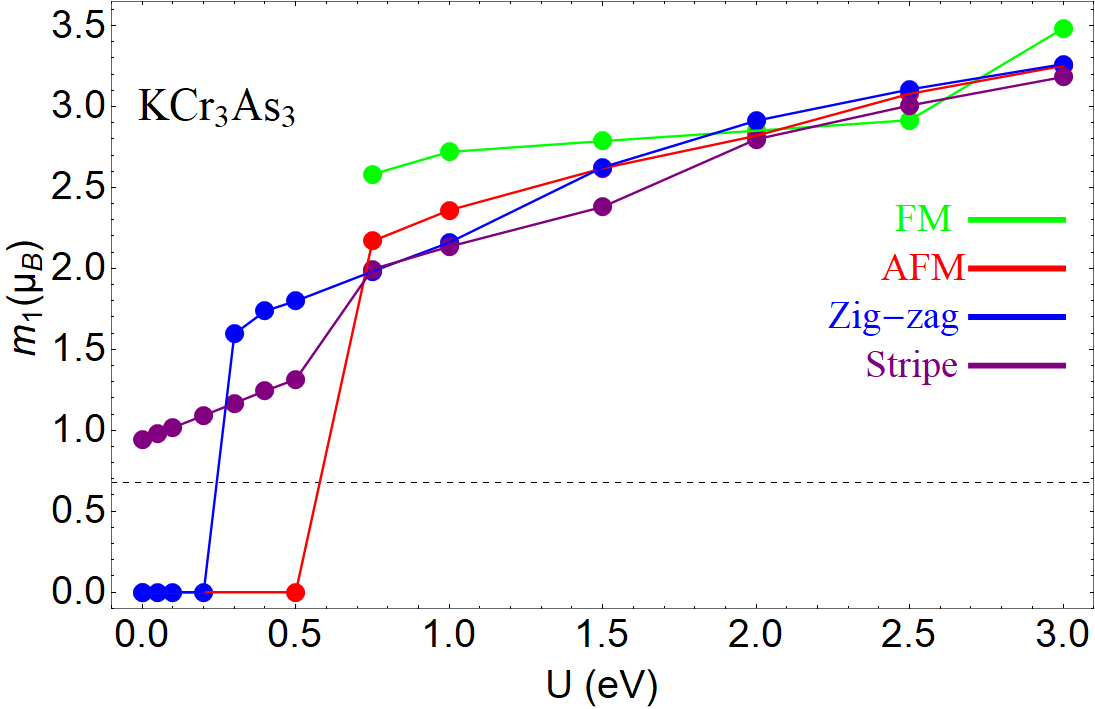}
	\includegraphics[width=\columnwidth,  angle=0]{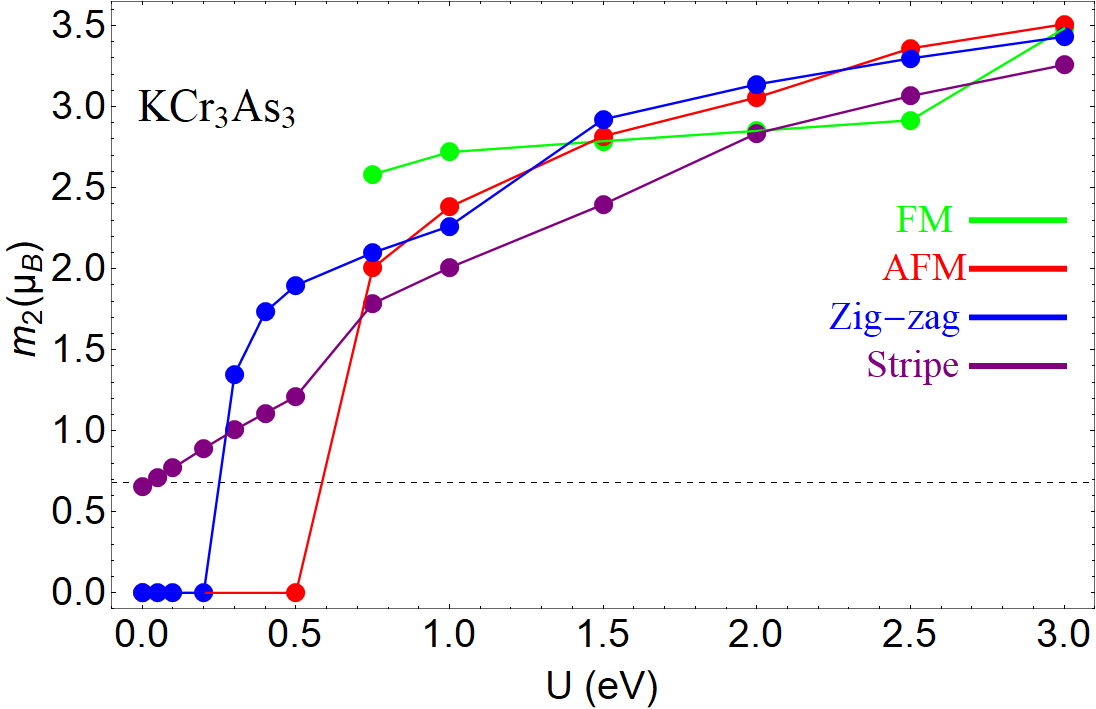}
	\caption{Magnetic moments of the Cr atoms for the configurations shown in Fig. \ref{configurations} as functions of the Coulomb interaction for the KCr$_{3}$As$_{3}$ compound. The experimental value is represented by the horizontal dashed line.}
	\label{magneticmoment_KCr3As3}
\end{figure*}

As far as the magnetic moments of the four inequivalent Cr sites are concerned, we may distinguish, as for the angle variations, a different behavior in the two regimes corresponding to values of $U$ approximately lower and higher than 1.5 eV, respectively (see Fig.~\ref{magneticmoment_K}).
For $U~\gtrsim~1.5\,$eV the magnetic moment at any Cr-site is quite large, assuming a value around 3 $\mu_B$ regardless of the magnetic configuration under consideration. We also note that with increasing $U$ these magnetic moments approach the maximum predicted value for Cr ions in the K$_{2}$Cr$_{3}$As$_{3}$ compound, that is 3.33 $\mu_B$. Indeed, since we have K$^{+1}$ and As$^0$ with empty Cr-4s levels, the valence of the chromium is 6.67 while the oxidation state is +2/3 in K$_{2}$Cr$_{3}$As$_{3}$.
On the other hand, at low values of $U$ the Cr ion magnetic moments  tend to vanish in the zig-zag as well as in the stripe phase, thus providing evidence of a non-magnetic ground state configuration.

\subsection{KCr$_3$As$_3$}

Let us now discuss the results obtained for KCr$_3$As$_3$. We will refer here to the same notation used in the previous subsection. 

We have plotted in Fig.~\ref{EC_KCr3As3} the energies of the various phases as functions of $U$, evaluated as in Fig.~\ref{EC_K} with respect to the energy of the stripe phase state. We find that the NM state is the ground state for $U\lesssim U_c=\,$0.3 eV, whereas for $U\gtrsim U_c$ the ground state is the stripe phase state, as in the case of K$_2$Cr$_3$As$_3$.

Differently from K$_2$Cr$_3$As$_3$, where we have one layer of K$_1$ and one of K$_3$, in KCr$_3$As$_3$ the two layers at $z=0$ and $z=0.5$ are equivalent, and so are the vertex angles and the corresponding Cr atoms belonging to the adjacent planes.
We plot in Fig.~\ref{angles_KC3As3} the value of the vertex angle of the triangle located at $z=0$. In the stripe configuration this angle does not vary significantly with increasing $U$, implying a quite robust stable configuration against variations of the Coulomb repulsion. Looking at the local magnetic moments (Fig.~\ref{magneticmoment_KCr3As3}), we find, as in K$_2$Cr$_3$As$_3$, two distinct behaviors in the regimes of low and high values of $U$. Indeed, for $U\lesssim 0.5\,$eV the zig-zag and the AFM phases exhibit a vanishing local magnetic moment, whereas in the stripe configuration a finite value is found also at $U=0$. On the other hand, for large $U$ the magnetic moments approach the maximum value predicted for chromium ions in the KCr$_{3}$As$_{3}$ compound, that is 3.67 $\mu_B$, due to a valence of 6.33 electrons and a corresponding oxidation state equal to +1/3.

Since it has been experimentally found that KCr$_3$As$_3$ has a finite magnetic moment, approximately equal to 0.68 $\mu_B$ \cite{Bao15b}, we can say that our results are consistent with a picture of KCr$_3$As$_3$ being in a moderately correlated ground state of the stripe type where $U$ is approximately equal to 0.3 eV and the predicted value of the magnetic moment is finite, though slightly larger ($\sim 1.1 \mu_B$) than the above-mentioned experimental one.
In addition, the degree of distortion of the triangles forming the DWSN is in this state predicted to be 72 degrees, as one can see from Fig.~\ref{angles_KC3As3}.

\section{Heisenberg model for magnetic exchanges inside the chain}

The DFT analysis described above demonstrated that the lowest energy magnetic configurations within the DWSN are stable already for low values of the on-site Coulomb repulsion. Also, the magnetic susceptibility measurements suggest dominantly antiferromagnetic interactions between local moments at Cr sites at least in K$_2$Cr$_3$As$_3$~\cite{Wu15}. We thus may combine those results to develop a magnetic exchange model which is able to describe the ground state magnetic configurations within the Cr triangles. In particular, to get these results we will consider collinear as well as non-collinear magnetic configurations. As already well-established for CrAs~\cite{Matsuda18,Sen19}, since the Cr$_3$As$_3$-chain-K-based materials may be considered as itinerant magnets, the mapping on an Heisenberg model is on the verge of the applicability, so that the calculation we will present below should be considered as merely qualitative. Indeed, strictly speaking, the Heisenberg model is only justified for systems with localized moments, such as insulators or rare earth
elements, but not for materials where the itinerant electrons are responsible for magnetism. 
Nevertheless, it may work reasonably well also for some systems belonging to this class of materials, a remarkable
example being bcc-Fe showing spin-spiral configurations~\cite{Singer11}. Moreover, we point out that the Heisenberg model we put forth to capture the magnetic picture outlined in the previous Section, retains all the symmetries of the lattice site configurations of the systems under investigation. Therefore, the calculation we will present below refers to a mapping of the DFT results into a Heisenberg model and assuming only planar configurations, as made in the previous Section.

Preliminarily, we have calculated the magnetocrystalline anisotropy $K$ obtaining $K=0.01\,$meV per magnetic atom for values of $U$ up to 3\,eV. We are thus confident that our assumption about planar configurations may result reasonable. Hence, we assume two independent Heisenberg coupling constants in each plane as well as two different magnetic coupling constants between the planes. We stress that, as for the first neighbours of CrAs~\cite{Autieri17,Matsuda18,Sen19}, also the first neighbours of K$_2$Cr$_3$As$_3$~\cite{Jiang15,Wu15} are antiferromagnetically coupled.
Therefore, the intrachain magnetic exchanges are antiferromagnetic, the magnetic moments lying in the $ab$ plane.

Taking into account these assumptions and considering the magnetic configurations as represented in Fig.~\ref{model}, we will adopt the following Heisenberg Hamiltonian:

\begin{figure}[]
	\includegraphics[width=\columnwidth,  angle=0]{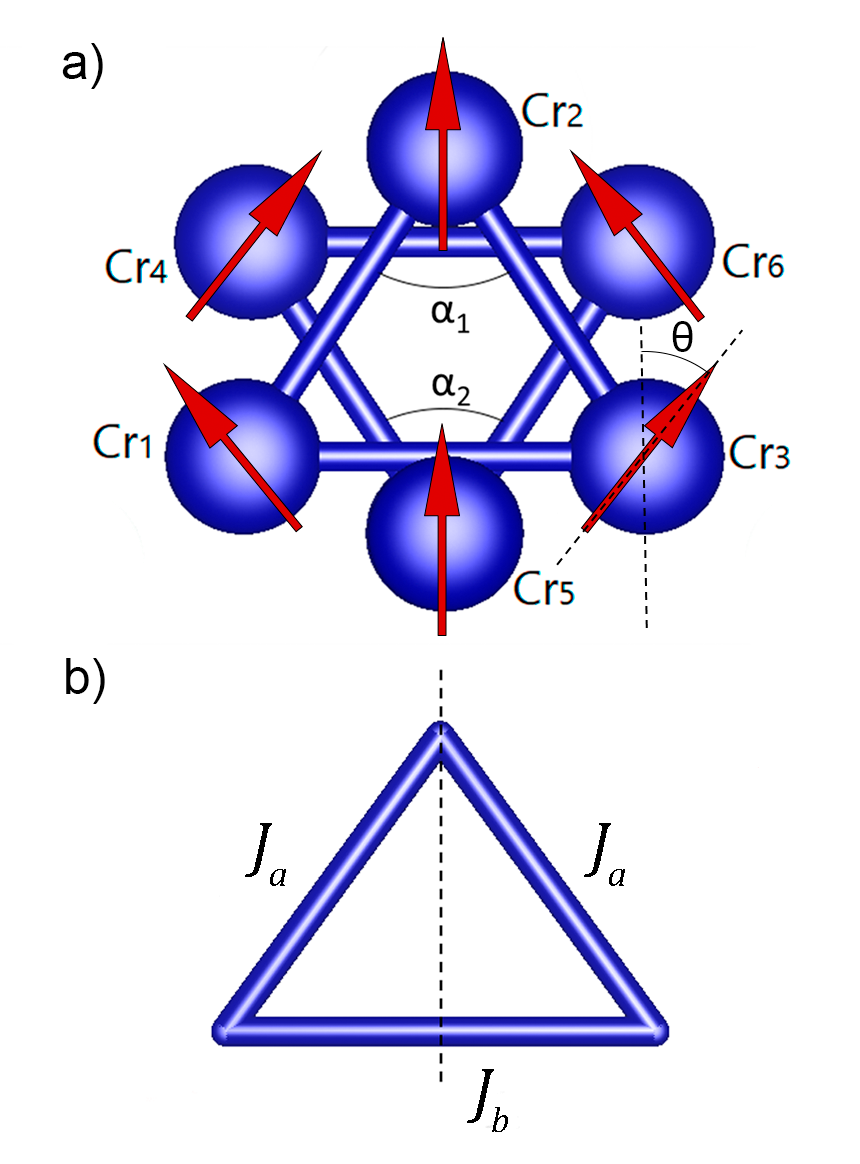}
	\caption{a) Magnetic arrangement of the Cr-ion magnetic moments in the $z$=0 and $z$=0.5 planes taken into account in the theoretical model defined by the Hamiltonian ~\eqref{Heisenberg}. b) Exchange coupling constants: $J_{b}$ is the coupling between the atoms at the basis of the triangle, while $J_{a}$ is between the atom at the basis and 
		the atom at the apex. The dashed line is the mirror axis. }
	\label{model}
\end{figure}

\begin{equation}\label{Heisenberg}
H=\sum_{<i,j,\mu\leq \nu>} J_{i,j}^{\mu,\nu} S_{i}^{\mu}{\cdot}S_{j}^{\nu}\quad .
\end{equation}
Here the sum is over pairs of adjacent spins in the $z=0$ and/or $z=0.5$ planes, and $\mu,\nu\in \{0,1\}$. If $\mu,\nu=0$ the spins are both in the plane located at $z=0$, if $\mu,\nu=1$ they both lie in the plane at $z=0.5$, if $\mu\ne\nu$ the two spins are in different planes.
We assume the mirror symmetry with respect to the $y$-axis and a rotation of the spin at the basis of the triangles by an angle $\theta$ as shown in Fig.~\ref{model}(a).  
In a classical picture, the total energy for a single triangle, expressed as a function of the angle $\theta$ between the spins, is given by:
\begin{equation}\label{TOTEN_1}
E^{\mu}(\theta)=S^2\left [ 2J_{a}^{\mu,\mu}\cos(\theta) +J_{b}^{\mu,\mu}\cos(2\theta)\right ] \quad .
\end{equation}
Here the angle $\theta$ ranges from 0 to $\pi$, $J_{b}^{\mu,\mu}$ ($J_{a}^{\mu,\mu}$) is the magnetic coupling between the ions Cr$_1$ and Cr$_3$ (Cr$_1$ and Cr$_2$ or Cr$_3$ and Cr$_2$) when the ions are located at $z=0$ ($\mu=0$), and between the ions Cr$_4$ and Cr$_6$ (Cr$_4$ and Cr$_5$ or Cr$_6$ and Cr$_5$) when they are located at $z=0.5$ ($\mu=1$). Summarizing, $J_{b}^{0,0}=J_{1,3}^{0,0}$ and $J_{b}^{1,1}=J_{4,6}^{1,1}$, while $J_{a}^{0,0}=J_{1,2}^{0,0}=J_{3,2}^{0,0}$ and $J_{a}^{1,1}=J_{4,5}^{1,1}=J_{6,5}^{1,1}$.

The coupling energy between the triangles is given by
\begin{equation}\label{TOTEN_2}
E^{0,1}(\theta)=S^2\left [J_{d}^{0,1}\cos(\theta) + J_{e}^{0,1}\cos(2\theta)\right ]\, ,
\end{equation}
where $J_{d}^{0,1}$ and $J_{e}^{0,1}$ are the inter-plane coupling constants. Specifically,  $J_{d}^{0,1}=J_{2,4}^{0,1}+J_{2,6}^{0,1}+J_{1,5}^{0,1}+J_{3,5}^{0,1}$=4$J_{2,4}^{0,1}$ and $J_{e}^{0,1}=J_{1,4}^{0,1}+J_{3,6}^{0,1}$=2$J_{1,4}^{0,1}$. 
From the available experimental data, we know that the couplings are all antiferromagnetic~\cite{Jiang15,Wu15} and the inter-plane coupling constants are numerically smaller than the in-plane ones, 
so that $J_{a}^{\mu,\mu},J_{b}^{\mu,\mu} > J_{d}^{\mu,\nu},J_{e}^{\mu,\nu} > 0$. This was also verified in the Appendix B.

Taking into account all the intra- and inter-plane interactions,
we finally get the total energy 
$E_{\rm tot}(\theta)=E^{0}(\theta)+E^{1}(\theta)+E^{0,1}(\theta)$ as a function of $\theta$:
\begin{equation}\label{TOTEN}
E_{\rm tot}(\theta)=S^2 J_{\rm tot}\left [ \cos(\theta)+ x \cos(2 \theta) \right ]\, .
\end{equation}
Here, $J_{\rm tot}=2(J_{a}^{0,0}+J_{a}^{1,1})+J_{d}^{0,1}$ and $x$=$(J_{b}^{0,0}+J_{b}^{1,1}+J_{3}^{0,1})/J_{\rm tot}$.
Looking at Fig.~\ref{model}, we note that when $\alpha_1=\alpha_2=60^{\circ}$, namely when the triangles are equilateral, we have $J_{a}^{\mu,\mu}=J_{b}^{\mu,\mu}$, when $\alpha_{k}<60^{\circ}$, $J_{a}^{\mu,\mu}<J_{b}^{\mu,\mu}$, and when $\alpha_{k}>60^{\circ}$, $J_{a}^{\mu,\mu}>J_{b}^{\mu,\mu}$. Finally, regardless of the values of  $\alpha_1$ and $\alpha_2$, we have $x\le 1$.

Studying the parametric Eq.~(\ref{TOTEN}), we find that the extreme points are given by $\theta_{ext}=0$, $\pi$, $ arccos(-\frac{1}{4x})$.
The value $\theta_{ext}=0$ gives the ferromagnetic solution (Fig.~\ref{configurations}(a)) and corresponds to the highest-energy ground state. The second one, $\theta_{ext}=\pi$, corresponds to the collinear stripe configuration depicted in Fig.~\ref{configurations}(c). The third one, $\theta_{ext}$=$ arccos(-\frac{1}{4x})$, gives a non collinear phase corresponding to the so-called in-out solution previously investigated in literature in the case of equilateral triangles~\cite{Jiang15,Wu15}.  We emphasize that for low values of the constant $x$, i.e. for $x \le 0.25$, the lowest energy solution corresponds to the collinear ferrimagnetic state, that we have previously called stripe configuration, while for $x>0.25$ the ground state becomes the non-collinear solution.  

Thus, this simple Heisenberg model calculation supports the DFT results presented in the previous Section. Nonetheless, we stress once again that the compounds we are dealing with are metallic so that the outcomes presented in this section should be considered as merely qualitative.


\section{Comparative analysis of magnetic properties of Cr$_3$As$_3$-chain-based K-materials}

The compound KCr$_3$As$_3$ does not superconduct, and exhibits a cluster-spin-glass behavior below $T_N=5\,$K \cite{Bao15b}. We point out that this value is close to the superconducting critical temperature of K$_2$Cr$_3$As$_3$ \cite{Bao15}, so that the energy scales involved in the formation of the two phases are essentially the same.  At higher temperatures, it shows a Curie-Weiss behavior, indicating the presence of non-vanishing local moments \cite{Bao15b}. Owing to the close relation with K$_2$Cr$_3$As$_3$, the presence of local-moment magnetism in KCr$_3$As$_3$ suggests that the existence of finite moments on chromium sites is detrimental to the development of a superconducting phase.

\begin{figure}[]
	\includegraphics[width=\columnwidth,  angle=0]{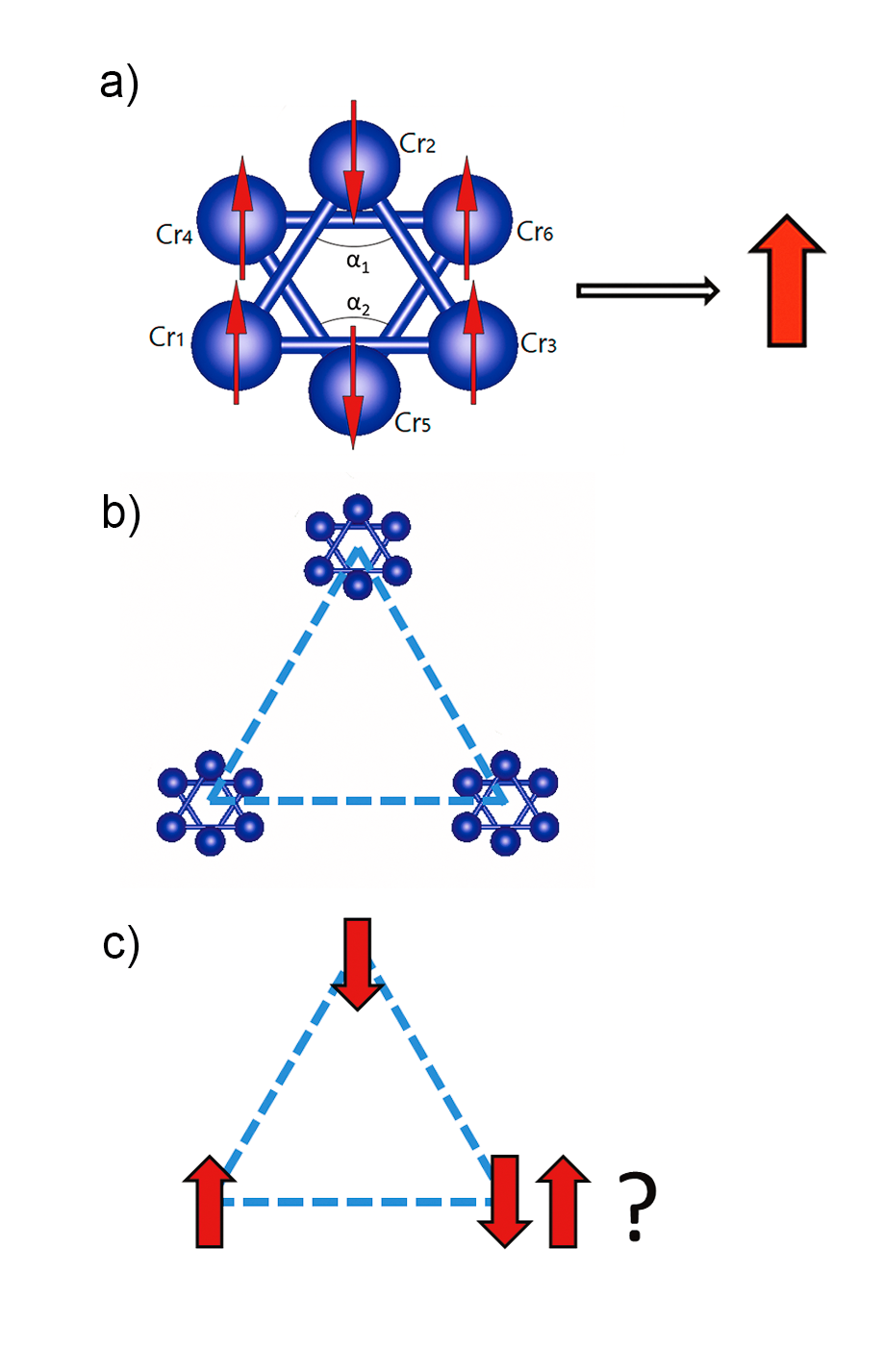}
	\caption{ a) A net magnetic moment can be attributed to a single chain, because the ground state within the DWSN predicted by our calculations is a collinear stripe configuration.  b) Interaction between the single chains. c) Schematic representation of the spin-glass behavior originating from geometrical frustrations between the  Cr-twisted tubes.}
\label{intertube}	
\end{figure}

The spin-glass-like state in KCr$_3$As$_3$ seems to be associated with the geometrical frustrations in the  Cr-twisted tubes. Moreover, the very appearance of a superconducting phase in K$_2$Cr$_3$As$_3$, which is indeed undetected in KCr$_3$As$_3$, may be  related to the existence of unequal Cr sites \cite{Cao17,Wu15,Wu15b}, as recent high-pressure studies \cite{Wang16} suggest. 
In order to explore the magnetic instabilities of KCr$_3$As$_3$ and K$_2$Cr$_3$As$_3$, we extend our DFT study considering different intertube magnetic interactions, possibly arising in the two different space groups.
For this purpose, we summarize here some of the main outcomes of the previous sections. Our DFT results predict that, for both compounds, the ground state is non-magnetic up to a critical value $U_c$ of the Coulomb repulsion, approximately equal to $0.3\,$eV for KCr$_3$As$_3$ and $0.4\,$eV for K$_2$Cr$_3$As$_3$. We thus have that in the interval between these two values KCr$_3$As$_3$ is magnetic while K$_2$Cr$_3$As$_3$ is non magnetic. 
Above $U_c$, a collinear stripe configuration is predicted within the DWSN, which allows to attribute a net magnetic moment to each chain. 
Such scheme is depicted in Fig.~\ref{intertube}, where we show the macrospin $\Uparrow$, which is associated to each chain (panel a)). Single interchain magnetic interactions then couple spins of neighboring chains, which are located at the sites of a triangular lattice. Without interchain magnetic coupling, the system can only show magnetic order in 1D, this being difficult to achieve because of the Mermin-Wagner theorem \cite{Gelfert01,Noce06}. With the presence of the interchain magnetic coupling, the system is not quasi-one-dimensional anymore and a magnetic order can more easily develop. This crucial interchain magnetic coupling can be ferromagnetic or antiferromagnetic. We are going to show that in the case of ferromagnetic interchain coupling, the system is ferrimagnetic with a net magnetic moment, while in the case of antiferromagnetic interchain coupling the system is likely to exhibit a spin-glass phase.

We proceed by looking at the energy of some possible interchain magnetic configurations. In this calculation we have used the PBEsol approximation for a supercell of two chains, for which we have considered the interchain FM, the interchain AFM and the NM phases. In the FM configuration, the chains have collinear magnetic moments $\Uparrow$, oriented in the same direction; in the AFM case they have collinear moments, which are antiparallel on the neighboring chains, while in the NM configuration all magnetic moments are zero.  Due to the estimated values of $U_c$, in this study we limited to analyze the relevant cases of $U=0, 0.3, 0.5\,$eV. 
The results reported in Tab.~\ref{Tabamongchains1} indicate that for K$_2$Cr$_3$As$_3$ the nonmagnetic phase is the more stable one up to $U=0.3\,$eV, whereas above this value the coupling between the chains is weakly ferromagnetic. On the other hand, in the case of KCr$_3$As$_3$ we observe that for $U\lesssim 0.3\,$eV the ground state is non-magnetic and for $U\gtrsim 0.3\,$eV the ferromagnetic coupling and the antiferromagnetic one are almost degenerate.
We used also different approximations, namely the LDA and the METAGGA, and we obtained similar results, confirming in the KCr$_3$As$_3$ the competition between the ferromagnetic and the antiferromagnetic interactions. In Fig.~\ref{intertube}(c) we give a schematic representation of this competition, in terms of geometrically frustrated AFM coupling between magnetic moments located at the sites of a triangular lattice. Frustration is known to be a prerequisite for the spin glass behavior, leading to an equal probability for the spins of being aligned or anti-aligned, and thus preventing long-range magnetic order. 
We propose that the magnetic coupling between the chains is antiferromagnetic for the KCr$_3$As$_3$ driving the system towards the spin-glass behavior and determining the scale of the critical temperature. Since the magnetic coupling is related to long-range hopping parameters, this coupling is expected to be relatively small, as also shown from the data of Table \ref{Tabamongchains1}.
The experimental critical temperature T$_N$=5 K  is compatible with the critical temperature of other frustrated systems where the relevant magnetic coupling is not between the first-neighbor sites.
\cite{Ivanov16}

\noindent 
\begin{table}[t!]
\begin{centering}
\begin{tabular}{|c|c|c|c|c|c|c|}
\hline 
\multicolumn{7}{|c|}{PBEsol} \tabularnewline
\hline 
 & \multicolumn{3}{c|}{K$_{2}$Cr$_{3}$As$_{3}$} & \multicolumn{3}{c|}{KCr$_{3}$As$_{3}$}\tabularnewline
\hline 
U (eV) & 0 & 0.3 & 0.5 & 0 & 0.3 & 0.5 \tabularnewline
\hline  
NM  & 0  & 0 & 19.92 & 0   & 3.75 & 19.33 \tabularnewline
\hline 
FM  & conv. to NM  & 6.92 & 0 & 12.08  & 0 & 0 \tabularnewline
\hline 
AFM  & conv. to NM  & 10.83 & 3.83 & 12.17  & 0.58 & 1.08 \tabularnewline
\hline  
\end{tabular}
\par\end{centering}
\caption{ Energy (meV) in the PBEsol approximation, measured with respect to the ground state one, corresponding to various inter-chain couplings, assuming the intra-chain collinear configuration. NM, FM and AFM indicate non-magnetic, ferromagnetic and antiferromagnetic inter-chain arrangements, respectively. Inside the chain the configuration is the $\uparrow\uparrow\downarrow$/$\uparrow\uparrow\downarrow$.  }
\label{Tabamongchains1} 
\end{table}

\section{Discussion and conclusions}

\begin{figure}[]
	\includegraphics[width=\columnwidth,  angle=0]{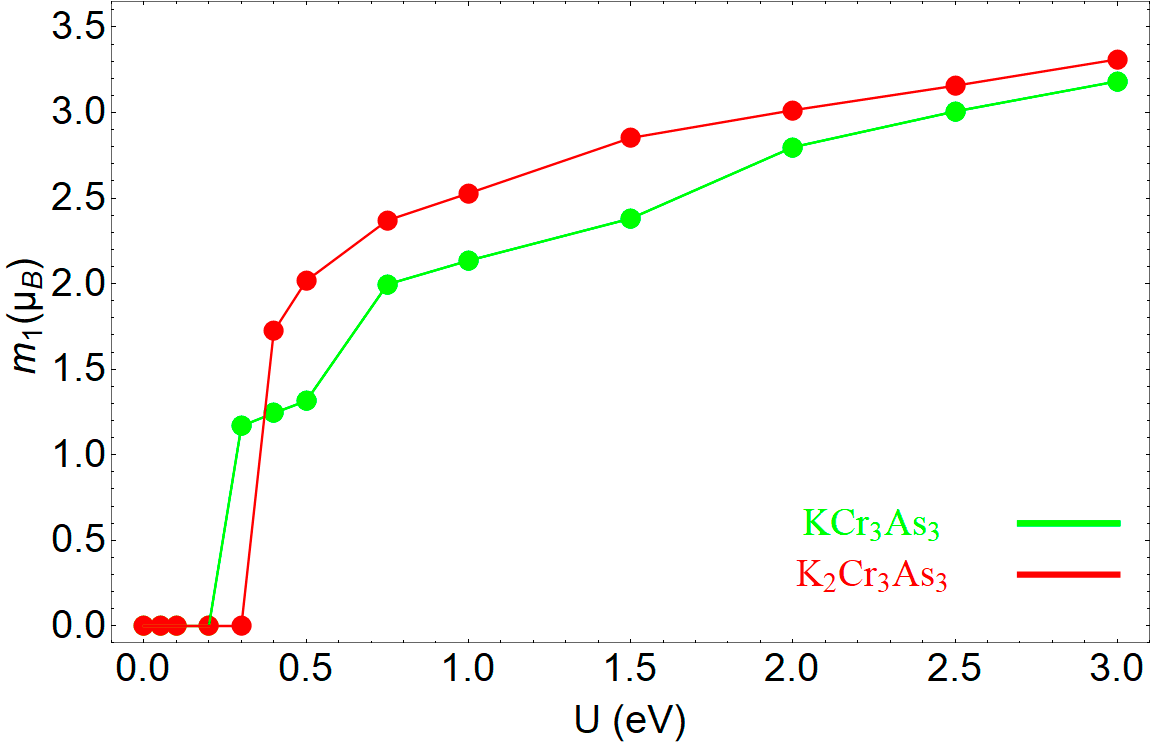}
	\caption{Magnetic moment of the Cr$_{1}$ ion for the KCr$_{3}$As$_{3}$ and K$_2$Cr$_{3}$As$_{3}$ compounds in the ground state.}
	\label{comparison}
\end{figure}

It is known that in  Cr$_3$As$_3$-chain-based K-materials there is a strong interplay between spin and lattice degrees of freedom, this issue being supported by the monotonic reduction of the superconducting critical temperature with the lattice expansion produced by the substitution of the K ion with larger alkaline ions. Nevertheless, the magnetic properties of different Cr$_3$As$_3$-chain-based K-materials are rather distinct, suggesting again the important role played by the crystal structure.

In this paper we performed first-principles calculations to investigate the magnetic phases which are compatible with the orthorhombic distortion of the CrAs tubes in this family of compounds. In Fig.~\ref{comparison} we summarize the comparison between K$_2$Cr$_3$As$_3$ and KCr$_3$As$_3$, by plotting the local magnetic moment at the representative Cr$_1$ site in the ground state, as a function of the Coulomb repulsion $U$. We observe that in both cases a non-vanishing magnetic moment can be attributed to the Cr sites above $U_c$, which slightly differs for the two compounds. Above this threshold, our DFT solution converges to a  collinear stripe phase, which gives rise to a nonzero magnetization within the chain.

In the previous section we have predicted that, in the energy window 0.3 eV$\,\lesssim U \lesssim \,$0.5 eV where KCr$_3$As$_3$ is in such collinear stripe phase and the value of the magnetic moment is sufficiently close to the experimentally detected one, the ferromagnetic and antiferromagnetic couplings among the chains are close in energy. We conjecture that this result may be at the origin of a spin-glass behavior driven by the geometric frustration of the magnetic coupling among different chains, consistently with the experimental results reported by Bao et al. \cite{Bao15b}. 
We also propose that the magnetic coupling between the chains in KCr$_3$As$_3$ determines the low critical temperature experimentally observed for the spin-glass phase \cite{Bao15,Bao15b}.

In the same region of the parameter space, we predict K$_2$Cr$_3$As$_3$ to be non-magnetic but on the verge of magnetism, sustaining interchain ferromagnetic spin fluctuations while the intrachain spin fluctuations are antiferromagnetic.
These findings are also in agreement with previous experiments. In particular, it was reported that the imaginary part of the bare electron susceptibility of K$_2$Cr$_3$As$_3$ shows large peaks at the point $\Gamma$, suggesting the presence of large ferromagnetic spin fluctuations in the compound \cite{Jiang15}, that are robust and persist in the case of non-equilateral triangles. We note that these results suggest an analogy with the ruthenates \cite{Mackenzie03,Cuoco06,Cuoco06b,Forte10,Autieri12,Malvestuto13,Autieri14,Granata16}. Indeed, the compound Ca$_2$RuO$_4$ is magnetic with an antiferromagnetic coupling between the Ru-ions,\cite{Porter2018} whereas Sr$_2$RuO$_4$ in on the verge of magnetism, exhibiting in the magnetic phase a ferromagnetic coupling among the Ru-ions. 

In this respect, we underline the importance of the investigation of the effects of a compressive strain applied orthogonally to the basis of the triangles in the chain of K$_2$Cr$_3$As$_3$ \cite{Cuono20b}. In a study of these effects, we have found an increase of the apical angle $\alpha$ which, by virtue of the above mentioned interplay between structural properties and magnetism, enforces the stability of the magnetism \cite{Cuono20b}. Based on these arguments, it has also been demonstrated that, through a compressive strain one can tune the superconducting K$_2$Cr$_3$As$_3$ toward a ferrimagnetic system, providing a playground to investigate the interplay between magnetism and superconductivity \cite{Galluzzi20} in this class of compounds \cite{Cuono20b}.



We finally remark that our picture of K$_2$Cr$_3$As$_3$ as a moderately correlated system is consistent with
ARPES measurements \cite{Watson17} showing that the overall bandwidth of the Cr 3$d$ bands as well as the Fermi velocities are in agreement with the DFT results. A similar conclusion can also be inferred from a recent investigation of the band structure of this compound \cite{Cuono19c}, where it has been shown that a good agreement with the results available in the literature \cite{Jiang15,Watson17} can be obtained for low values of the electronic correlation.
We are then confident that the interesting features outlined in this paper in the moderately correlated regime are quite robust.

As a final remark, we notice that the proximity to a stripe ferrimagnetic phase within the chains may turn out to be relevant to obtain indications on the mechanism driving the superconducting phase in K$_2$Cr$_3$As$_3$. Studies in this direction are planned for the next future. 


\begin{figure}
	\centering
	\includegraphics[width=\columnwidth, angle=0]{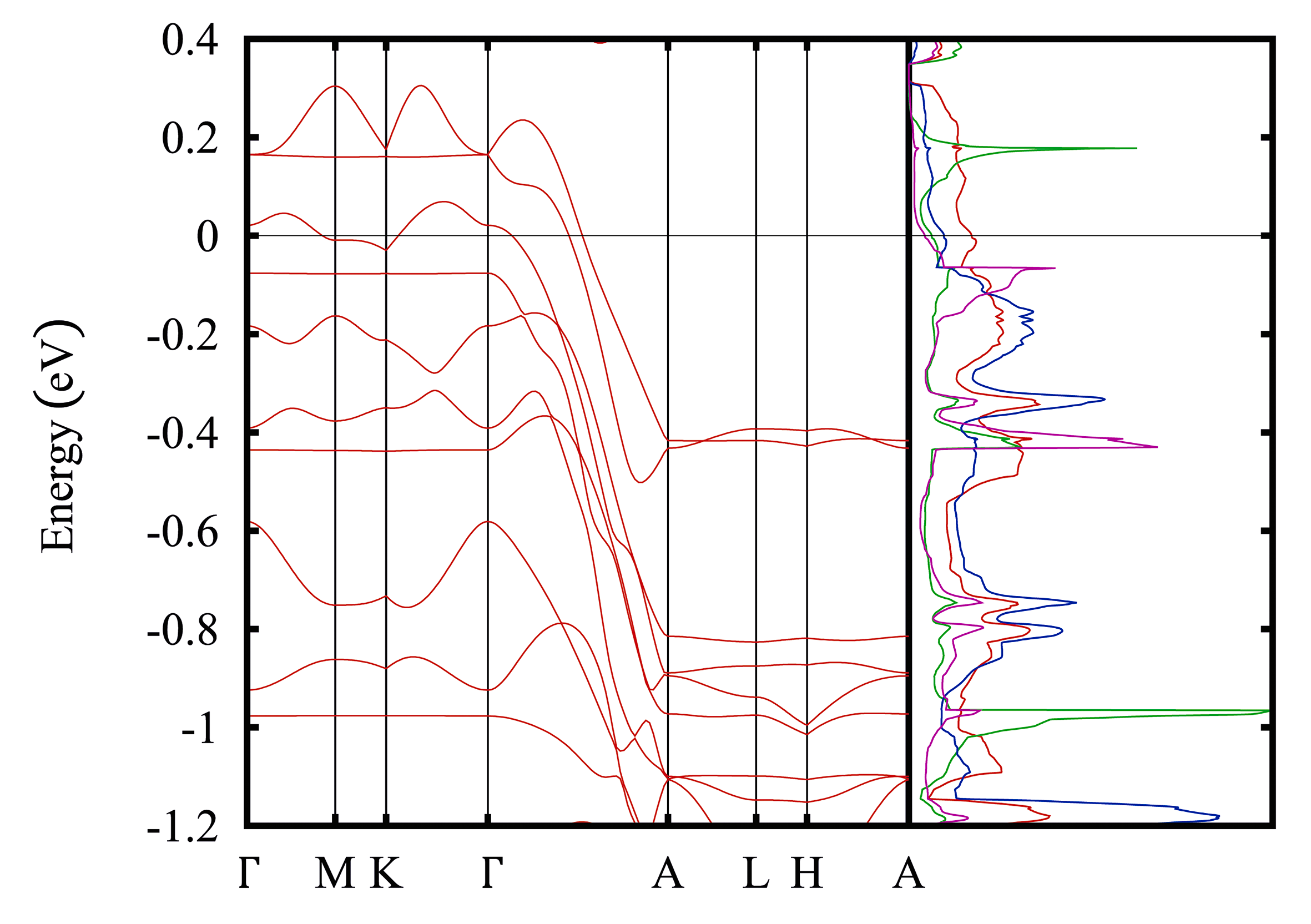}
	\caption{Band structure of the K$_2$Cr$_3$As$_3$ near the Fermi level for U = 0.3 eV (left), together with the corresponding orbitally-resolved partial densities of states (right). Red and green lines refer to the Cr$_1$ atom at the basis of the triangle and denote the DOSs projected onto orbitals symmetric with respect to the basal plane d$_{xy}$, d$_{x^2-y^2}$ and d$_{z^2}$ and onto orbitals antisymmetric with respect to the basal plane d$_{yz}$ and d$_{xz}$, respectively. Blue and purple lines refer to the Cr$_2$ atom at the vertex of the triangle and denote the DOSs projected onto the symmetric and antisymmetric orbitals, respectively.
	}
	\label{BandDOSK2}
\end{figure}

\section*{Acknowledgments}
The authors acknowledge A. Galluzzi and M. Polichetti for useful discussions. 
G. C. acknowledges financial support from ''Fondazione
Angelo Della Riccia''.
The work is supported by the Foundation for Polish Science through the International Research
Agendas program co-financed by the European Union within the Smart Growth Operational Programme. 
X. M. was sponsored by the National Natural Science Foundation of China (No. 11864008).
We acknowledge the access to the computing facilities of the
Interdisciplinary Center of Modeling at the University of
Warsaw, Grant No. G73-23 and G75-10. 
We acknowledge the CINECA award under the
ISCRA initiatives IsC69 “MAINTOP” and IsC76 "MEPBI"
Grant, for the availability of high-performance computing
resources and support.

\section*{Appendix A: Electronic properties}

In this appendix we analyze the electronic properties of K$_2$Cr$_3$As$_3$ and KCr$_3$As$_3$ in the distorted case. 
As explained in Section III, we fix for the Coulomb repulsion the value $U=0.3\,$eV, for which KCr$_3$As$_3$ is magnetic while K$_2$Cr$_3$As$_3$ is non-magnetic.
%
In Fig.~\ref{BandDOSK2} we report the band structure and the orbitally-resolved partial densities of states for the $d$ orbitals of K$_2$Cr$_3$As$_3$ near the Fermi level. 
We consider the partial densities of states for the $d$ orbitals symmetric respect to the basal plane d$_{xy}$, d$_{x^2-y^2}$ and d$_{z^2}$ and the antisymmetric with respect to the basal plane d$_{yz}$ and d$_{xz}$.
We make this distinction because we know from the undistorted case \cite{Jiang15,Cuono19c} that symmetric orbitals with respect to the basal plane d$_{xy}$, d$_{x^2-y^2}$ and d$_{z^2}$ have a greater spectral weight at the Fermi level, while the antisymmetric ones have a relevant weight few eV far from the Fermi level.
In the distorted structure the atoms at $z=0$ and $z=0.5$ are closer than in the undistorted case, suggesting that the hybridization between the orbitals at different planes is larger, both in the symmetric and in the antisymmetric orbital manifolds, as well as between symmetric and antisymmetric orbitals. For simplicity, we report only the results for the $d$ orbitals because, as for the undistorted case, they encompass the largest weight at the Fermi level.
We have shown in Ref. \onlinecite{Cuono19c} that, apart from a slight increase of the energy bandwidth, corresponding to the chromium states
being pushed away from the Fermi level, electronic correlations on chromium orbitals barely affect the energy
spectrum. Therefore, we can say that the band structure obtained for the chosen value of the Coulomb repulsion would look similar to the band structure at vanishing $U$. The reasonable agreement between our results and ARPES measurements \cite{Watson17} indicates that these materials could be considered as moderately correlated compounds that can be satisfactorily described considering small values of the Coulomb repulsion.

\begin{figure}
	\centering
	\includegraphics[width=\columnwidth, angle=0]{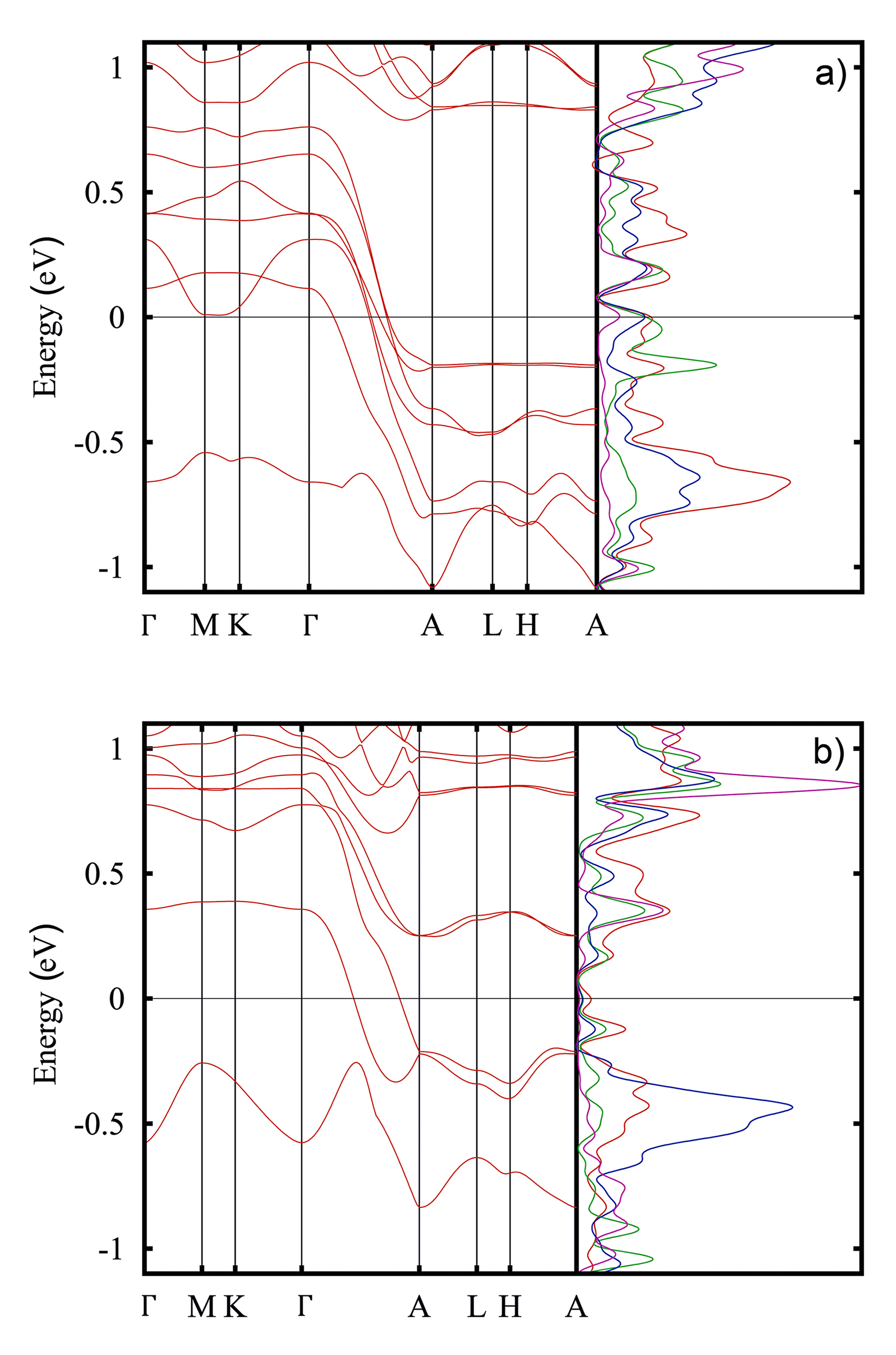}
	\caption{a) Band structure of the KCr$_3$As$_3$ near the Fermi level for U = 0.3 eV (lef), together with the corresponding orbitally-resolved partial densities of states in the spin-up channel (right). The same colors of Fig. \ref{BandDOSK2} are used for the local DOSs. b) The same as in the top panel for the spin-down channel.
	}
	\label{BandDOSK1}
\end{figure}

Since, for $U=0.3\,$eV the ground state of KCr$_3$As$_3$ is magnetic, we report in Fig.~\ref{BandDOSK1} the band structure and the orbitally-resolved partial densities of states for the $d$ orbitals near the Fermi level, in the spin-up and in the spin-down channel respectively. The results are in good agreement with the ones quoted in Ref. \onlinecite{Xing19}, where it is shown that the prominent three-dimensional (3D) Fermi surface in the undistorted structure is not present in the distorted case. Indeed, the band structure along the in-plane high-symmetry lines of the Brillouin zone is significantly different from the undistorted case. Actually, the bands are more flat and the symmetry properties that allow to connect the bands along some lines are removed by the distortion. For completeness, we noted that in our case, at $U=0.3\,$eV, the system is ferrimagnetic, differently from Ref. \onlinecite{Xing19} where the system is studied in the non-magnetic phase.
\\

\begin{figure}
  \includegraphics[scale=0.45]{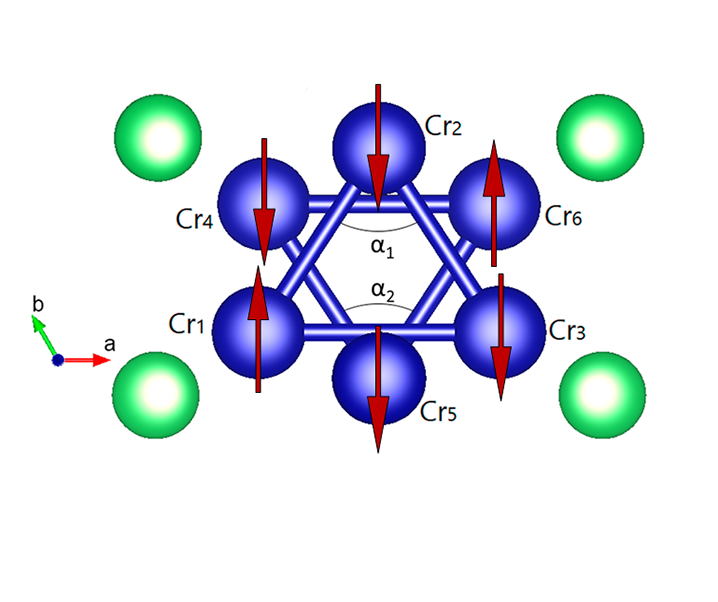}
	\caption{Arrangements of the Cr-spin of the additional magnetic configuration investigated to obtain the magnetic exchanges. The Cr atoms at the basis of the triangles have opposite moment directions.
	}
	\label{newconfiguration}
\end{figure}

\begin{figure}
  \includegraphics[width=\columnwidth, angle=0]{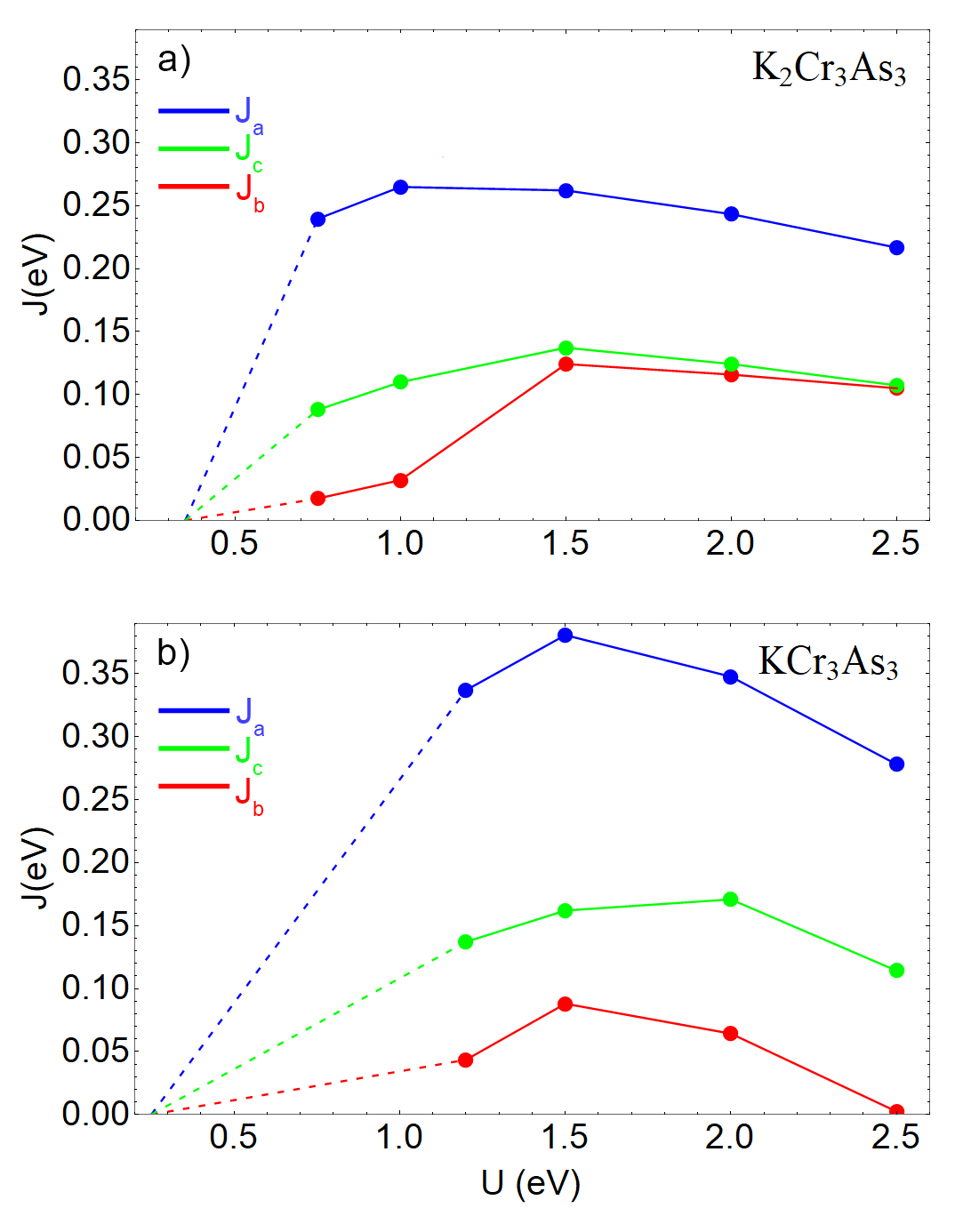}
	\caption{a) Magnetic exchanges for the K$_2$Cr$_3$As$_3$ as a function of the Coulomb repulsion U. 
	J$_a$, J$_b$ and J$_c$ are three magnetic exchanges described in the text.
	The dashed line indicates the extrapolation of the couplings in the region where the mapping is not possible.
	b) The same as in panel a) for  KCr$_3$As$_3$.
	}
	\label{Jvalues_U}
\end{figure}

\section*{Appendix B: Magnetic exchanges}

In this appendix, we map the DFT results on the Heisenberg model providing an estimation of the exchange couplings and explaining the limitations of the method for this class of materials.
The mapping of the DFT results on the Heisenberg model relies on the calculation of the energy of several magnetic configurations. 
The reliability of this method largely depends on the values of the magnetic moments that should be constant in these different configurations. 
As shown in Figs.~\ref{magneticmoment_K} and \ref{magneticmoment_KCr3As3}, while the magnetic moments are approximately constant in the various magnetic phases for $U\gtrsim1.5\,$eV, they strongly differ for $U<1.5\,$eV. However,  
our results show that it is still possible to perform the mapping to the Heisenberg model for 0.75\,eV $\lesssim U <$ 1.5\,eV in the case of K$_2$Cr$_3$As$_3$, and for 1.2 eV $\lesssim U<$ 1.5\,eV in the case of KCr$_3$As$_3$, though
we are on the verge of the applicability of the mapping for these materials.
Since we know that the magnetic coupling should go to zero for $U$ approximately lower than 0.4\,eV, in the case of K$_2$Cr$_3$As$_3$ we may roughly extrapolate the value of $J$ from $U=0.75\,$eV to $U=0.30\,$eV, as shown by the dotted line in Fig.~\ref{Jvalues_U}a). 
Similarly, in the case of KCr$_3$As$_3$ we are able to reliably calculate the magnetic couplings up to $U=1.2\,$eV, and, as in the previous case, we extrapolate an estimation of the exchange couplings considering that they should go to zero for $U$ lower than 0.3 \,eV (dotted line in Fig.~\ref{Jvalues_U}b)).
We note that in order to assess the first-neighbor magnetic coupling, we assume that the two interlayer ones are equal, namely $J_{2,4}^{0,1}$=$J_{1,4}^{0,1}$ (the same notation of Section V is here adopted), and we have calculated the energy also considering the magnetic configuration plotted in Fig.~\ref{newconfiguration}. We see that in this additional configuration, the spins of the Cr atoms at the basis of the triangles have opposite directions.
For completeness, we point out that we have excluded from our evaluation the ferromagnetic phase configuration since it can be obtained from a suitable linear combination of the other phases.
We indicate with $J_a$ the magnetic coupling between the Cr at the apex and the Cr at the basis of the isosceles triangles, while $J_b$ is the magnetic coupling between the two Cr-atoms of the basis. $J_c$ is the inter-plane coupling constant,  
$J_c$=$J_{2,4}^{0,1}$=$J_{1,4}^{0,1}$ using the notation of the Section V.
The results we obtain for the magnetic exchanges as  $U$ is varied, are reported in Fig.~\ref{Jvalues_U}a) for K$_2$Cr$_3$As$_3$ and in Fig.~\ref{Jvalues_U}b) for KCr$_3$As$_3$.
From an inspection of these figures, we infer that the magnetic exchanges for KCr$_3$As$_3$ are on average larger than the magnetic exchanges for the K$_2$Cr$_3$As$_3$. Also, all the first-neighbor magnetic exchanges are antiferromagnetic, for every value of $U$, including the case $U=2\,$eV previously reported for the undistorted case \cite{Wu15}. In both cases, the in-plane magnetic coupling $J_a$ is larger than $J_b$, stabilizing the collinear magnetic configuration, as reported in the main text. Furthermore, the interlayer magnetic coupling $J_c$ is larger than the previously calculated magnetic couplings \cite{Wu15}, and we attribute this difference to the effect of the distortions that reduces the interlayer Cr-Cr bonds and, consequentially, increases the coupling between the magnetic moments.

Finally, as far as the estimation of the critical temperature is concerned, we note that the systems investigated here are quasi-one-dimensional with frustrated magnetism. Accordingly, a mean-field approach fails to provide a correct estimation of $T_c$. This would rather require the inclusion in the calculation of the inter-chain exchange interactions, but this is beyond the scope of the present paper. 


\end{document}